\documentstyle[aps,preprint,epsfig]{revtex}

\newcommand{\bq}{\begin{equation}}
\newcommand{\eq}{\end{equation}}
\newcommand{\bqa}{\begin{eqnarray}}
\newcommand{\eqa}{\end{eqnarray}}
\newcommand{\nn}{\nonumber \\}

\begin{document}
\draft 
\title{
Absence of time-reversal symmetry breaking in association with the order 
parameter of Cooper pair in high $T_c$ superconductivity
}
\author{Sung-Sik Lee, Ki-Seok Kim, Jae-Hyun Eom and Sung-Ho Suck Salk$^1$}
\address{Department of Physics, Pohang University of Science and Technology,\\
Pohang, Kyoungbuk, Korea 790-784\\
$^1$ Korea Institute of Advanced Studies, Seoul 130-012, Korea\\}
\date{\today}

\maketitle

\begin{abstract}
As an extension of our previous work on the holon pairing instability in the t-J 
Hamiltonian [Phys. Rev. B {\bf 66}, 054427 (2002)], 
we examine 
the orbital symmetries of holon pairing order parameters in high $T_c$ 
superconductivity by examining the energy poles of t-matrix.
We find that both $s$- and $d$-wave symmetries in holon pair order parameter 
occur at low lying energy states corresponding to the 
higher energy 
poles of t-matrix while only the $s$-wave symmetry appears at the lowest energy 
pole and that this results in the $d$-wave symmetry 
in the 
Cooper pair order which is a composite of the holon pair of $s$-wave symmetry 
and the spinon pair of $d$-wave symmetry below $T_c$.
Finally we demonstrate that there exists no time-reversal symmetry breaking in 
association with the Cooper pair order parameter.
\end{abstract}
%\pacs{PACS numbers : 74.25.Bt, 74.25.Jb} 
%\begin{multicols}{2}

\newpage
\section{INTRODUCTION}

At present the  Cooper pair order parameter of $d_{x^2 - y^2}$-wave symmetry for 
high $T_c$ cuprates is generally 
accepted\cite{HARLINGEN_TSUEI}.
However, recent tunneling measurements on $Y Ba_2 Cu_3 O_{7-3}$ compound show 
that pairing symmetry changes from a pure $d_{x^2 - 
y^2}$-wave 
symmetry to a mixture of $d$- and $s$- wave symmetries with the change of hole 
concentration\cite{DAGAN,YEH,SHARONI} or with the 
variation of  
applied magnetic field\cite{DAGAN}.
Cooper pair order parameters of complex mixing such as $d_{x^2 - y^2} + id_{xy}$ 
and $d_{x^2 - y^2} + is$ break the time-reversal 
symmetry\cite{DAGAN,SHARONI}. 
There have been some theoretical attempts to explain the observed thermal 
conductivity\cite{KRISHANA} in high $T_c$ cuprates by 
resorting to the 
time-reversal symmetry breaking of the Cooper pair wave 
function\cite{LAUGHLIN,GHOSH}.
On the other hand, the signatures for time-reversal symmetry breaking have also 
been found in the angle resolved photoemission 
spectroscopy\cite{KAMINSKI}, neutron scattering\cite{MOOK} and $\mu$-SR 
measurements\cite{SONIER} below the pseudogap temperature 
$T^*$.
Earlier, Varma\cite{VARMA} proposed that the fourfold pattern of circulating 
current inside the $Cu O_2$ unit cell breaks the 
time-reversal 
symmetry.
Chakravarty et al.\cite{CHAKRAVARTY} suggested that the time-reversal symmetry 
is broken by the d-density wave order which involves 
the 
circulating current around the $Cu$-$O$ bond.

Recently we\cite{LEE1} reported the superconducting phase diagram by applying 
the Bethe-Salpeter equation to our earlier U(1) and 
SU(2) 
holon-pair boson theory\cite{LEE2} of t-J Hamiltonian and obtained an arch shape 
structure of superconducting transition temperature 
in 
agreement with experiments.
The $d_{x^2 - y^2}$-wave symmetry of Cooper pair can arise as a composite of the 
$s$-wave symmetry of the holon pair and the $d_{x^2 
- 
y^2}$-wave symmetry of spinon pair. 
In the previous t-matrix study we briefly reported a discussion on the $d$-wave 
symmetry of the Cooper pair order parameter at and 
below $T_c$. 
Here we present a detailed study on how the orbital symmetry of the holon 
pairing order parameters varies as a function of 
excitation energy 
below and above $T_c$. 
It is of great interest to examine whether the widely used t-J Hamiltonian can 
intrinsically predict the time reversal symmetry 
breaking by 
allowing a complex orbital  mixing.
The objectives of the present paper are two-fold.
First we discuss the variation of the orbital symmetry for holon pair order 
parameter with excitation energy by examining the 
t-matrix pole.
Second we examine whether there exists the time reversal symmetry breaking  in 
association with the Cooper pair order parameter 
based on the 
U(1) and SU(2) slave-boson theories of the t-J Hamiltonian.

\section{U(1) and SU(2) slave-boson reprensentations of t-matrix based on t-J 
Hamiltonian}
We write the t-J Hamiltonian, 
\begin{eqnarray}
H & = & -t\sum_{}(c_{i\sigma}^{\dagger}c_{j\sigma} + c.c.) + J\sum_{}({\bf 
S}_{i} \cdot {\bf S}_{j} - \frac{1}{4}n_{i}n_{j}),
\label{eq:tjmodel1}
\end{eqnarray}
where ${\bf S}_{i} \cdot {\bf S}_{j} - \frac{1}{4}n_{i}n_{j} =  -\frac{1}{2} ( 
c_{i \downarrow}^{\dagger}c_{j 
\uparrow}^{\dagger}-c_{i 
\uparrow}^{\dagger}c_{j \downarrow}^{\dagger}) (c_{j \uparrow}c_{i 
\downarrow}-c_{j\downarrow} c_{i\uparrow})$.
Here ${\bf S}_{i}$ is the electron spin operator at site $i$, ${\bf 
S}_{i}=\frac{1}{2}c_{i\alpha}^{\dagger} {\sigma}_{\alpha 
\beta}c_{i\beta}$ 
with ${\sigma}_{\alpha \beta}$, the Pauli spin matrix element and $n_i$, the 
electron number operator at site $i$, 
$n_i=c_{i\sigma}^{\dagger}c_{i\sigma}$.
In the slave-boson representation\cite{ZOU,KOTLIAR,FUKUYAMA,NAGAOSA,UBBENS,WEN} 
the electron annihilation operator $c_{\sigma}$ of 
spin $\sigma$ 
is written as a composite of spinon $f_{\sigma}$ (spin $1/2$ and charge $0$ 
object) and holon $b$ (spin $0$ and $+e$ object) 
operators.
That is, $c_{\sigma}  =  b^\dagger f_{\sigma}$ in the U(1) theory and 
$c_{\alpha}  = \frac{1}{\sqrt{2}} h^\dagger \psi_{\alpha}$ in 
the SU(2) 
theory with $\alpha=1,2$, where $f_{\sigma}$($b$) is the spinon(holon) 
annihilation operator in the U(1) theory, and $\psi_1=\left( 
\begin{array}{c} f_1 \\ f_2^\dagger \end{array} \right)$ and  $\psi_2 = \left( 
\begin{array}{c} f_2 \\ -f_1^\dagger \end{array} 
\right)$ and $h 
= \left( \begin{array}{c}  b_{1} \\ b_{2} \end{array} \right)$ are respectively 
the doublets of spinon and holon annihilation 
operators in the 
SU(2) theory.
In the slave-boson representation, the Heisenberg interaction term is written
$H_J  =   -\frac{J}{2} \sum_{} b_i b_j b_j^{\dagger}b_i^{\dagger} 
(f_{i\downarrow}^{\dagger}f_{j\uparrow }^{\dagger}-f_{i\uparrow}^{\dagger}
f_{j\downarrow}^{\dagger})(f_{j\uparrow}f_{i\downarrow}-f_{j\downarrow} 
f_{i\uparrow})$ 
for the U(1) theory and
$H_J = -\frac{J}{2} \sum_{} ( 1 + h_i^\dagger h_i) ( 1 +  h_j^{\dagger}h_j ) 
(f_{i2}^{\dagger}f_{j1}^{\dagger}-f_{i1}^ 
{\dagger}f_{j2}^{\dagger})(f_{j1}f_{i2}-f_{j2} f_{i1})$
for the SU(2) theory.
In the above expressions the coupling between the charge (holon pair) and spin 
(spinon pair) degrees of freedom naturally arises 
from the 
composite nature of electron having both spin and charge.

After decomposition of the Heisenberg interaction term and proper Hubbard 
Stratonovich transformations, the effective mean field 
Hamiltonian of 
holon is derived to be\cite{LEE2}, for the U(1) theory
(see Appendix A for a derivation),
\bqa
H_{t-J,U(1)}^{b} & = &  -t \sum_{} \chi_{ij}^* b_{i}^{\dagger} b_{j} + c.c. \nn
& - &  \frac{J}{2} \sum_{} |\Delta^f_{ij}|^2 b_{i}^{\dagger} b_{j}^\dagger b_j 
b_i
- \mu \sum_{i}  b_{i}^{\dagger}b_{i} ,
\label{eq:u1_model}
\eqa
where $\chi_{ij}= < f_{i\sigma}^{\dagger}f_{j\sigma} + \frac{4t}{J_p} 
b_{i}^{\dagger}b_{j}>$ is the hopping order parameter and $ 
\Delta^f_{ij} 
= < f_{j\uparrow} f_{i\downarrow}-f_{j\downarrow}f_{i\uparrow} >$, the spinon 
pairing order parameter, and for the SU(2) slave-boson 
theory,
(see Appendix B for a derivation),
\bqa
\lefteqn{H_{t-J,SU(2)}^{b}  =  -\frac{t}{2} \sum_{} h_{i}^{\dagger} U_{ij} h_{j} 
+ c.c. }\nn
&& -   \frac{J}{2} \sum_{,\alpha,\beta} |\Delta^f_{ij}|^2 b_{i\alpha}^{\dagger} 
b_{j\beta}^\dagger b_{j\beta} b_{i\alpha}
- \mu \sum_{i}  h_{i}^{\dagger}h_{i} ,
\label{eq:su2_model}
\eqa
where 
$ U_{i,j}  =  \left( \begin{array}{cc} \chi_{ij}^* & - \Delta^f_{ij} \\
			   - \Delta^{f*}_{ij} & -\chi_{ij} 
		\end{array} \right)$ is the order parameter matrix of hopping 
order, $ \chi_{ij} $ and spinon pairing order, $ 
\Delta^f_{ij}$ 
with
$ \chi_{ij} = < f_{i\sigma}^{\dagger}f_{j\sigma} + \frac{2t}{J_p} 
(b_{i1}^{\dagger}b_{j1}-b_{j2}^{\dagger}b_{i2}) >$ and $ 
\Delta^f_{ij} = 
<f_{j1}f_{i2}-f_{j2} f_{i1}>$.

From the Bethe-Salpeter equation\cite{LEE1}, we obtain a matrix equation for the 
t-matrix, for the U(1) slave-boson theory 
(see Appendix C for a detailed derivation),
\bqa
\sum_{ {\bf k}^{''}} \left( \delta_{{\bf k}^{'}, {\bf k}^{''}} - m_{{\bf k}^{'}, 
{\bf k}^{''}} ( {\bf q}, q_0 )  \right) t_{{\bf 
k}^{''}, {\bf 
k}}( {\bf q}, q_0 ) = v( {\bf k}^{'} -  {\bf k} ),
\label{eq:matrix_equation_u1}
\eqa
where 
\bq
m_{{\bf k}^{'}, {\bf k}^{''}} ( {\bf q}, q_0 ) =
\frac{1}{N} v( {\bf k}^{'} -  {\bf k}^{''} ) \frac{n(\epsilon({\bf k}^{''})) + 
e^{\beta \epsilon(-{\bf k}^{''}+ {\bf q}) } 
n(\epsilon(-{\bf 
k}^{''}+{\bf q}))}{ iq_0 - (\epsilon(-{\bf k}^{''}+{\bf q}) + \epsilon({\bf 
k}^{''})) },
\label{eq:kernel_u1}
\eq
$v( {\bf k} ) = -J | \Delta_f |^2 \gamma_{{\bf k}}$, the momentum space 
representation of holon-holon interaction with $\gamma_{{\bf 
k}} = ( 
\cos k_x + \cos k_y)$ and
$n(\epsilon) = \frac{1}{e^{\beta \epsilon} - 1}$, the boson distribution 
function. 
Similarly, we obtain, for the SU(2) slave-boson theory
(see Appendix D for a derivation),
\bqa
\lefteqn{\sum_{ {\bf k}^{''}, \alpha^{''}, \beta^{''} } \left( \delta_{{\bf 
k}^{'}, {\bf k}^{''}} \delta_{\alpha^{'} \alpha^{''}} 
\delta_{\beta^{'} \beta^{''}} - m^{\alpha^{'} \beta^{'} \alpha^{''} \beta^{''} 
}_{{\bf k}^{'}, {\bf k}^{''}} ( {\bf q}, q_0 )  
\right) t^{ 
\alpha^{''} \beta^{''} \alpha \beta }_{{\bf k}^{''}, {\bf k}}( {\bf q}, q_0 ) 
}\nn
&& = v( {\bf k}^{'} -  {\bf k} ) \delta_{\alpha^{'} \alpha} \delta_{\beta^{'} 
\beta}, \hspace{4cm}
\label{eq:matrix_equation_su2}
\eqa
where 
\bqa
m^{\alpha^{'} \beta^{'} \alpha^{''} \beta^{''} }_{{\bf k}^{'}, {\bf k}^{''}} ( 
{\bf q}, q_0 ) & \equiv &
\frac{1}{N} \sum_{\alpha^{'}_1 \beta^{'}_1} v( {\bf k}^{'} -  {\bf k}^{''} ) 
\frac{n(E_{\alpha^{'}_1}({\bf k}^{''})) + e^{\beta 
E_{\beta^{'}_1}(-{\bf k}^{''}+{\bf q})} n(E_{\beta^{'}_1}(-{\bf k}^{''}+{\bf 
q}))}{ iq_0 - (E_{\alpha^{'}_1}({\bf k}^{''}) + 
E_{\beta^{'}_1}(-{\bf k}^{''}+ {\bf q} )) } \times  \nn
&& U_{\alpha^{'}\alpha^{'}_1}({\bf k}^{''}) U_{\beta^{'}\beta^{'}_1}(- {\bf 
k}^{''} + {\bf q}) U^\dagger_{\alpha^{'}_1 
\alpha^{''}}({\bf 
k}^{''}) U^\dagger_{\beta^{'}_1 \beta^{''} }(- {\bf k}^{''} + {\bf q}). 
\label{eq:kernel_su2}
\eqa
Here $E_{1}( {\bf k} ) = E_{{\bf k}} - \mu$ and $E_{2}( {\bf k} ) = -E_{{\bf k}} 
- \mu$ are the energy dispersions of upper and 
lower bands of 
holons with $E_{{\bf k}} =  t \sqrt{ (\chi \gamma_{{\bf k}})^2 + ( \Delta_f 
\varphi_{{\bf k}})^2 }$ and $\varphi_{{\bf k}} = ( \cos 
k_x - \cos 
k_y)$.
$U_{\alpha \beta}({\bf k}) = \left( \begin{array}{cc} u_{{\bf k}} & -v_{{\bf k}} 
\\ v_{{\bf k}} & u_{{\bf k}} \end{array} \right)$ 
is the 
unitary matrix which diagonalizes the one-body holon Hamiltonian with $ u_{{\bf 
k}} = \frac{1}{\sqrt{2}}\sqrt{1-\frac{t\chi 
\gamma_{{\bf 
k}}}{E_{{\bf k}}}}$ and $ v_{{\bf k}} = 
\frac{sgn(\varphi_k)}{\sqrt{2}}\sqrt{1+\frac{t\chi \gamma_{{\bf k}}}{E_{{\bf 
k}}}}$.
(Here sgn($\varphi_k$) denotes the sign of $(\cos k_x - \cos k_y)$.) 
It is noted that the t-matrices, $t_{{\bf k}^{'}, {\bf k}}( {\bf q}, q_0 )$ and 
$t^{ \alpha^{'} \beta^{'} \alpha \beta }_{{\bf 
k}^{'}, {\bf k}}( 
{\bf q}, q_0 )$ for the U(1) and SU(2) theories respectively are independent of 
the Matsubara frequencies $k^{'}_0$ and $k_0$, owing 
to the 
consideration of instantaneous holon-holon interaction, $v( {\bf k}^{'} -  {\bf k})$. 

The t-matrices are numerically evaluated from the use of 
Eqs.(\ref{eq:matrix_equation_u1}) and (\ref{eq:matrix_equation_su2}) at 
each 
temperature and hole doping concentration.
The hopping and spinon pairing order parameters in Eqs.(\ref{eq:u1_model}) and 
(\ref{eq:su2_model}) are the saddle point values 
evaluated from 
the usual partition functions involving the functional integrals of slave-boson 
representation.\cite{LEE2}
Using the matrix equations (\ref{eq:matrix_equation_u1}) and 
(\ref{eq:matrix_equation_su2}), the poles of the t-matrices are 
searched for as a 
function of energy $q_0$ with ${\bf q} = 0$, i.e., the zero momentum of the 
holon pair by using
\bqa
\sum_{{\bf k^{'}}} t_{{\bf k}, {\bf k}^{'}}( q_0, {\bf q}= 0 ) W(q_0, {\bf 
k^{'}}) = \lambda W(q_0, {\bf k}),
\label{eq:eigenequation_u1}
\eqa
for the U(1) theory, where $W (q_0, {\bf k})$ is the eigenvector and $\lambda$, 
the eigenvalue and 
\bqa
\sum_{ {\bf k^{'}}, \alpha^{'}, \beta^{'}}  t^{ \alpha \beta \alpha^{'} 
\beta^{'} }_{{\bf k}, {\bf k}^{'}} ( q_0, {\bf q}= 0 )  
W_{\alpha^{'} 
\beta^{'}}(q_0, {\bf k}^{'}) = \lambda W_{\alpha \beta}(q_0, {\bf k}),
\label{eq:eigenequation_su2}
\eqa
for the SU(2) theory, where $W_{\alpha \beta}(q_0, {\bf k})$ is the eigenvector 
associated with the SU(2) isospin channels $\alpha$ 
and 
$\beta$($\alpha = 1, 2$ and $\beta = 1, 2$ ) and $\lambda$, the corresponding 
eigenvalue.
The divergence of the eigenvalue at a given energy $q_0$ signifies the pole of 
the t-matrix at the energy.
The value of pole $q_0$ corresponds to the energy of holon pair corresponding to 
the eigenvector $W(q_0,{\bf k})$. 

 For each eigenvector we compute the contribution of the $s$-, $p_x$-, $p_y$- 
and $d_{x^2-y^2}$-orbitals.
The orbital symmetry is  determined from numerical fitting to the computed 
result of the eigenvector,
\bq
W(q_0, {\bf k}) = 
a_s ( \cos k_x + \cos k_y )
+a_{p_x}  \sin k_x 
+a_{p_y}  \sin k_y 
+a_d ( \cos k_x - \cos k_y )
\eq
for the U(1) theory and
\bq
W_{\alpha \beta} (q_0, {\bf k}) = 
a_{s, \alpha \beta} ( \cos k_x + \cos k_y )
+ a_{p_x, \alpha \beta}  \sin k_x 
+a_{p_y, \alpha \beta}  \sin k_y 
+a_{d, \alpha \beta} ( \cos k_x - \cos k_y )
\eq
for the SU(2) theory.
Here, $a_l$ represents the weight of the $l$-th partial wave of holon pair in 
the U(1) theory with orbital angular momentum of 
$l=s$, $p_x$, 
$p_y$ and $d$.
For the SU(2) theory, $a_{l, \alpha \beta}$ represents the weight of the $l$-th 
partial wave in the $b_{\alpha}$ and $b_{\beta}$ 
holon pairing 
channel with $\alpha,\beta =1$, $2$. 
The poles and eigenvectors are obtained for three different hole concentrations 
: underdoped, optimally doped and overdoped cases 
with $J/t = 
0.3$ and $N = 10 \times 10$ square  lattice.

\section{RESULTS}

In Fig. 1, we summarize our results for the energy poles and corresponding 
eigenvectors of the t-matrix.
Above the on-set temperature $T_c$, there is no phase coherent holon pair bound 
state. 
As temperature is lowered to a temperature below $T_c$ from a temperature above 
$T_c$, the lowest energy pole changes from a 
positive to 
negative sign, that is, a phase coherent bound state with negative energy 
occurs\cite{LEE1}.
The predicted bound state has the $s$-wave pairing symmetry.
As shown in Fig. 1, the energies of the higher lying pairing states remains to 
be positive (unbound) even below $T_c$.

In Table 1, we show the weight of each orbital for the eigenvector of the U(1) 
t-matrix at the lowest energy at temperature slightly 
above the 
on-set temperature ($T = T_c + 10^{-3}t$ with $t$, the hoping integral).
The eigenvector of the lowest energy corresponds to the stable channel of holon 
pairing below $T_c$\cite{LEE1}.
As shown in Table 1, stable  holon pairing occurs in the pure $s$-wave channel. 
The contributions of the higher partial wave ($p$- and $d$-wave) symmetries 
virtually vanishes;
they are found to be less than $10^{-13}$ times smaller than that of the 
$s$-wave contribution.
The lowest lying holon pairing state remains to be the pure $s$-wave (having no 
mixing with other orbitals) with decreasing 
temperature from a 
temperature above $T_c$.
It is now clear from this prediction that the Cooper pairing occurs with the 
pure $d_{x^2 - y^2}$ symmetry resulting from the 
composite nature 
of the $d_{x^2 - y^2}$-wave symmetry of spinon pairing and the $s$-wave symmetry 
of the holon pairing\cite{COOPER_U1}.

With the SU(2) t-matrix, there exist two eigenvectors corresponding to two 
bosons $b_1$ and $b_2$.
The weight of each orbital for the first eigenvector is shown in Table 2 and the 
second, in Table 3.
As shown in Table 2, the first eigenvector is of pure $s$-wave pairing symmetry 
with no phase difference between the $b_1$-$b_1$ and 
$b_2$-$b_2$ 
pair scattering channel.
As shown in Table 3, the second eigenvector shows a dominance of the $s$-wave 
pairing symmetry with the phase difference of $\pi$ 
between the 
$b_1$-$b_1$ and $b_2$-$b_2$ channels and a negligibly small contribution from 
the $d$-wave symmetry in the $b_1$-$b_2$ channel.
The weights of other orbital  contributions are negligibly small as shown in the 
Tables.
It can be readily checked that only the second eigenvector in Table 3 leads to 
non-vanishing Cooper pair order parameter, while the 
first one in 
Table 2 results in vanishing Cooper pair order parameter\cite{COOPER_SU2}.
The SU(2) theory differs from the U(1) theory in that the phase fluctuations of 
order parameters are taken into account by allowing 
the 
$b_2$-boson\cite{WEN}. 
There are two eigenvectors of t-matrix associated with the same energy pole in 
SU(2).
However, there is only one stable $s$-wave channel of holon pairing with no 
other orbital mixing which leads to the non-vanishing 
Cooper pair 
order parameter for both the U(1) and SU(2) theories.
It is now clear that the composition of the $s$-wave channel of the holon pair 
and the $d$-wave channel of the spinon pair leads to 
the pure 
$d_{x^2 - y^2}$-wave symmetry of the Cooper pair order 
parameter\cite{HARLINGEN_TSUEI}.
Thus there exists no time-reversal symmetry breaking allowing no complex orbital 
mixing.

For the higher energy poles, the $s$-, $p_x$-, $p_y$- and $d_{x^2 - y^2}$-wave 
states are predicted to occur as shown in Tables 4 
and 5.
Interestingly, the eigenvectors occur in a sequence of certain pattern as  the 
energy pole of the t-matrix increases both above and 
below $T_c$ 
as shown in Fig. 1 :
$s$-wave state occurs in the lowest energy;
the next four high lying energy states are of the $s$-, $p_x$-, $p_y$- and 
$d_{x^2 - y^2}$-wave symmetries with nearly the same 
energy (that is, 
near degenerate);
the next higher lying energy states are of $s$-, $p_x$- and $p_y$-wave 
symmetries with nearly the same energy and so on. 
Below we will discuss how the above sequential pattern arises in association 
with momenta available in the intermediate scattering 
states of 
holon pair .

First we analyze the locations of the poles.
The poles of the t-matrix occur at discrete energies for both bound and unbound  
states of the holon pairs for finite lattice.
With increasing lattice size ($L \rightarrow \infty$), the unbound states become 
continuum as the allowed momenta become continuum.
The energy of holon pair with momenta (${\bf k}$ and $-{\bf k}$) is the sum of 
kinetic energy and interaction energy.
The magnitude of the interaction energy ($V \sim J |\Delta_f|^2$) is relatively 
small as compared to that of the kinetic energy ($K 
\sim t\chi$) 
with the ratio of $V/K \leq 0.1$.
Thus the kinetic energy dominantly determines the energy of holon pair. 
Indeed, the predicted locations of poles in Tables 4 and 5 are close to the 
kinetic energy of holon pair with momenta ${\bf 
k}=(0,0), 
(2\pi/L,0), ( 2\pi/L, 2\pi/L)$ and $(2\pi/L,0)$\cite{ENERGY}.

 Here we pay attention to the intermediate scattering states of holon pair.
This will allow us to examine how an initial holon pair state with an $l$-th 
partial wave states ($l=s, p_x, p_y$ or $d$) can be 
scattered into 
intermediate states having momenta ${\bf k}_1$ and $-{\bf k}_1$.
The transition amplitude from the initial state to the intermediate state is, to 
first order, 
\bqa
T_{l \rightarrow \{ {\bf k}_1, -{\bf k}_1 \} } & = & \sum_{\bf k} v( {\bf k}_{1} 
- {\bf k} ) w_l({\bf k}) \nn
& = & -J|\Delta_f|^2  \sum_{\bf k} \left[ \cos (k_{1x}-k_x) + \cos (k_{1y}-k_y) 
\right] w_l({\bf k}),
\label{eq:nonvanishing_ch_u1}
\eqa
where $w_l({\bf k})$ is the eigenvector with an $l$-th partial wave (orbital).
The Feynmann diagram for this process is displayed in Fig. 2.

For the initial $s$-wave pairing state $w_l({\bf k}) = (\cos k_x + \cos k_y)$, 
the transition amplitude is obtained to be
\bqa
T_{s \rightarrow \{ {\bf k}_1, -{\bf k}_1 \}  }& = & -J|\Delta_f|^2  \sum_{\bf 
k} \left[ \cos (k_{1x}-k_x) + \cos (k_{1y}-k_y) 
\right] (\cos k_x 
+ \cos k_y) \nn
&=& -\frac{N}{2}J|\Delta_f|^2 (\cos k_{1x} + \cos k_{1y}).
\label{eq:nonvanishing_ch_u1_s}
\eqa
This amplitude does not vanish as long as the momentum of the intermediate state 
(${\bf k_1}$) satisfies the condition of $(\cos 
k_{1x} + \cos 
k_{1y}) \neq 0$. 
In other words, the $s$-wave state has a non-vanishing transition amplitude to 
the intermediate state with momenta 
${\bf k}_1$ and $-{\bf k}_1$ if $(\cos k_{1x} + \cos k_{1y}) \neq 0$.
Thus, the $s$-wave scattering channel allows poles at the intermediate state 
energies of holon pairs with momenta ${\bf k}_1$ and 
$-{\bf k}_1$ 
as long as $(\cos k_{1x} + \cos k_{1y}) \neq 0$.
For example, the $s$-wave eigenvector occurs at poles near $2\epsilon(0,0)$, 
$2\epsilon(2\pi/L,0)$, $2\epsilon(2\pi/L,2\pi/L)$ and 
$2\epsilon(4\pi/L,0)$ as shown in Table 4.
This is because $(\cos k_{1x} + \cos k_{1y})$ does not vanish for momenta, 
$(0,0)$, $(2\pi/L,0)$, $(2\pi/L,2\pi/L)$ and 
$(4\pi/L,0)$.
Similarly, one can readily understand why $p$- and $d$-wave scattering states 
(channels) occur only at certain energy 
levels\cite{PD}.
For higher angular momentum channels such as $w_l(k) = \cos 2k_x$ or $w_l(k) = 
\sin 2k_x$, the transition amplitude in Eq. 
(\ref{eq:nonvanishing_ch_u1_s}) identically vanishes.
Therefore, t-matrix can not allow poles in the scattering channels corresponding 
to these higher orbital angular momentum states;
only the $s$, $p_x$, $p_y$ and $d_{x^2 - y^2}$ state can occur.

The above analysis with the U(1) theory is readily applied to the SU(2) case to 
find the location of poles and the corresponding 
eigenvectors.
The only difference is that in SU(2) the number of eigenvectors for each pole is 
doubled owing to the presence of two degenerate 
bosons $b_1$ 
and $b_2$.
These two eigenvectors are found to be degenerate.
One is the eigenvector with no phase difference between the $b_1 - b_1$ and $b_2 
-b_2$ pairing channel.
The other is the eigenvector with the phase difference of $\pi$. 
Interestingly, there is a small component of $b_1 - b_2$ boson pairing for the 
case of the latter.
The weight of this off-diagonal pairing component is negligibly small in the 
lowest energy pole (e.g. $\sim 10^{-4}$ for 
$\omega=0.0097$ with 
$x=0.07$), but become appreciably large at higher energy poles (e.g. $\sim 0.2$ 
for $\omega/t=1.1236$ with $x=0.07$). 
On the other hand, the weight of the $b_1 - b_2$ boson pairing state decreases 
with the increase of hole concentration as shown in 
Table 5.
This is due to the decreasing trend of the spinon pairing order parameter 
$\Delta_f$ with increasing hole concentration.
It is of note that the off-diagonal element between $b_1$  and $b_2$ bosons (or 
vice versa) in the Hamiltonian is proportional to 
the spinon 
pairing order parameter in Eq.(\ref{eq:su2_model}).

It is also noted that there occurs not only $s$- and $d$- wave states, but also 
$p$- wave state for holon pairing.
The $p$- wave state is not allowed because it has the odd value of the orbital 
angular momentum and violate the symmetric waves.
With the inclusion of exchange channel only $s$- and $d$- wave states appear in 
association with eigenvectors in the t-matrix as 
shown in Fig. 
3.

\section{SUMMARY}
In the present study, we investigated the orbital contributions to both U(1) and 
SU(2) t-matrices both above and below $T_c$ and 
examined the 
lowest energy and higher energy poles respectively by applying the 
Bethe-Salpeter equation to the U(1) and SU(2) holon-pair boson 
theory.
We showed that at the lowest energy (below $T_c$) there is only one holon 
pairing bound state which gives rise to the non-vanishing 
Cooper pair 
order parameter for both the U(1) and SU(2) theories.
We found that the holon pairing  at and below $T_c$ is made of the pure $s$-wave 
with no complex orbital mixing and thus the Cooper 
pair of 
d-wave symmetry as a composite of the s-wave holon pair and the d-wave spinon 
pair. As a consequence there exits no time reversal 
symmetry 
breaking. 
Thus we argue that symmetry breaking is caused by a different origin since the 
complex orbital mixing such as $d_{x^2-y^2} + 
id_{xy}$ and  
$d_{x^2-y^2} + is$ did not appear.

\newpage
\renewcommand{\theequation}{A\arabic{equation}}
\setcounter{equation}{0}
\section*{APPENDIX A: Derivation of the U(1) holon Hamiltonian} 
The t-J Hamiltonian of interest is given by,
\begin{eqnarray}
H & = & -t\sum_{<i,j>}(c_{i\sigma}^{\dagger}c_{j\sigma} + c.c.) 
+ J\sum_{<i,j>}( {\bf S}_{i} \cdot {\bf S}_{j} - \frac{1}{4}n_{i}n_{j}) \nn 
&& -\mu \sum_i c_{i\sigma}^{\dagger} c_{i\sigma} 
\label{tjmodel}
\end{eqnarray}
and the Heisenberg interaction term is rewritten 
\bqa
H_J & = & J \sum_{<i,j>} ({\bf S}_{i} \cdot {\bf S}_{j} - \frac{1}{4}n_{i}n_{j}) 
\nn
& = & -\frac{J}{2} \sum_{<i,j>} ( 
c_{i\downarrow}^{\dagger}c_{j\uparrow}^{\dagger}-c_{i\uparrow}^{\dagger}c_{j\downarrow}^{\dagger}) 
(c_{j\uparrow}c_{i\downarrow}-c_{j\downarrow} c_{i\uparrow}).
\label{Heisen}
\eqa
Here $t$ is the hopping energy and ${\bf S}_{i}$, the electron spin operator at 
site $i$, ${\bf 
S}_{i}=\frac{1}{2}c_{i\alpha}^{\dagger} {\bbox \sigma}_{\alpha \beta}c_{i\beta}$ with ${\bbox \sigma}_{\alpha \beta}$, the 
Pauli spin matrix element. $n_i$ is the electron number 
operator at 
site $i$, $n_i=c_{i\sigma}^{\dagger}c_{i\sigma}$.
$\mu$ is the chemical potential.

In the U(1) slave-boson representation\cite{KOTLIAR,FUKUYAMA,NAGAOSA,UBBENS}, 
with single occupancy constraint at site $i$ the 
electron 
annihilation operator $c_{i \sigma}$ is taken as a composite operator of the 
spinon (neutrally charged fermion) annihilation 
operator $f_{i 
\sigma}$ and the holon (positively charged boson) creation operator 
$b_i^\dagger$, and thus, $c_{i \sigma} = f_{i \sigma} b_i^\dagger$.
Rigorously speaking, it should be noted that the expression $c_{i \sigma} = 
b_i^\dagger f_{i\sigma}$ is not precise since these 
operators belong 
to different Hilbert spaces and thus the equality sign here should be taken only 
as a symbol for mapping.
Using $c_{i \sigma} = f_{i \sigma} b_i^\dagger$ and introducing the Lagrange 
multiplier term (the last term in 
Eq.(\ref{tj_u1_slave})) to 
enforce single occupancy constraint, the t-J Hamiltonian is rewritten,
\bqa
&& H  =  -t\sum_{<i,j>}\left( (f_{i\sigma}^{\dagger}b_i)(b_j^\dagger 
f_{j\sigma}) + c.c. \right)  \nn
&&  -\frac{J}{2} \sum_{<i,j>} b_i b_j b_j^{\dagger}b_i^{\dagger} 
(f_{i\downarrow}^{\dagger}f_{j\uparrow }^{\dagger}-f_{i\uparrow}^{\dagger}f_{j\downarrow}^{\dagger})(f_{j\uparrow}f_{i\downarrow}-f_{j\downarrow} 
f_{i\uparrow}) \nn
&& - \mu \sum_i f_{i\sigma}^{\dagger}b_i f_{i\sigma} b_i^\dagger - i \sum_i 
\lambda_i ( b_i^\dagger b_i + f_{i\sigma}^\dagger 
f_{i\sigma} -1 ).
\label{tj_u1_slave} 
\eqa

The coupling between physical quantities $A$ and $B$ is decomposed into terms 
involving fluctuations of $A$, i.e., $(A-<A>)$ and 
$B$, i.e., 
$(B-<B>)$, separately uncorrelated mean field contribution of $<A>$ and $<B>$ 
and correlation between fluctuations of $A$ and $B$, 
that is, 
$(A-<A>)$ and  $(B-<B>)$ respectively;
$AB = <B> A + <A> B - <A> <B> + (A-<A>)(B-<B>)$.
Setting $A=b_i b_j b_j^\dagger b_i^\dagger$ for charge (holon) contribution and 
$B=(f_{i \downarrow}^\dagger f_{j \uparrow}^\dagger - f_{i \uparrow}^\dagger 
f_{j \downarrow}^\dagger ) ( f_{j \uparrow} f_{i \downarrow} - f_{j \downarrow} f_{i \uparrow} )$ for spin (spinon) contribution,
the Heisenberg coupling term can be decomposed into terms involving coupling 
between 
the charge and spin fluctuations separately, the mean field contributions and 
coupling (correlation) between fluctuations (charge 
and spin 
fluctuations).
Using such decomposition of the Heisenberg interaction term for 
Eq.(\ref{tj_u1_slave}),  we write the U(1) Hamiltonian
\bqa
H_{t-J}^{U(1)} & = & 
-t\sum_{<i,j>}(f_{i\sigma}^{\dagger}f_{j\sigma}b_{j}^{\dagger}b_{i} + c.c.) \nn
&& -\frac{J}{2}  \sum_{<i,j>} \Bigl[  \Bigl< 
(f_{i\downarrow}^{\dagger}f_{j\uparrow}^{\dagger}-f_{i\uparrow}^{\dagger}
f_{j\downarrow}^{\dagger})
 (f_{j\uparrow}f_{i\downarrow}-f_{j\downarrow} f_{i\uparrow}) 
 \Bigr> 
 b_i b_j b_j^{\dagger}b_i^{\dagger}  \nn
&& +  \Bigl< b_i b_j b_j^{\dagger}b_i^{\dagger} \Bigr> (f_{i\downarrow}^{\dagger}f_{j\uparrow}^{\dagger}-f_{i\uparrow}^ 
{\dagger}f_{j\downarrow}^{\dagger})
(f_{j\uparrow}f_{i\downarrow}-f_{j\downarrow} f_{i\uparrow}) \nn
& & -   \Bigl< b_i b_j b_j^{\dagger}b_i^{\dagger} \Bigr> 
\Bigl< 
(f_{i\downarrow}^{\dagger}f_{j\uparrow}^{\dagger}-f_{i\uparrow}^{\dagger}f_{j\downarrow}^{\dagger})
(f_{j\uparrow}f_{i\downarrow}-f_{j\downarrow} f_{i\uparrow}) 
\Bigr> \nn
&& + \Bigl( b_i b_j b_j^{\dagger}b_i^{\dagger} - \Bigl< b_i b_j 
b_j^{\dagger}b_i^{\dagger}\Bigr> \Bigr)   \nn
&& \times 
\Bigl( (f_{i\downarrow}^{\dagger}f_{j\uparrow}^{\dagger}-f_{i\uparrow}^ 
{\dagger}f_{j\downarrow}^{\dagger} ) 
( f_{j\uparrow}f_{i\downarrow}-f_{j\downarrow} f_{i\uparrow}) 
- \Bigl<(f_{i\downarrow}^{\dagger}f_{j\uparrow}^{\dagger}-f_{i\uparrow}^ 
{\dagger}f_{j\downarrow}^{\dagger})
(f_{j\uparrow}f_{i\downarrow}-f_{j\downarrow} f_{i\uparrow})\Bigr> \Bigr) \Bigr] 
\nn
&& - \mu \sum_{i} f_{i\sigma}^{\dagger} f_{i\sigma} ( 1 + b_i^\dagger b_i )
- i\sum_{i} \lambda_{i}(f_{i\sigma}^{\dagger}f_{i\sigma}+b_{i}^{\dagger}b_{i} 
-1).
\label{u1_tjmodel0}
\end{eqnarray}

Noting that $[b_i, b_j^{\dagger}] = \delta_{ij}$ for boson, the intersite charge 
(holon) interaction term (the second term) in 
Eq.(\ref{u1_tjmodel0}) is rewritten,
\bqa
&& -\frac{J}{2} 
\Bigl< (f_{i\downarrow}^{\dagger}f_{j\uparrow}^{\dagger}-f_{i\uparrow}^ 
{\dagger}f_{j\downarrow}^{\dagger})(f_{j\uparrow}f_{i\downarrow}-f_{j\downarrow} 
f_{i\uparrow}) \Bigr> 
b_i b_j b_j^{\dagger}b_i^{\dagger} \nn
&& = -\frac{J}{2} <|\Delta_{ij}^f|>^2 \left( 1 + b_i^\dagger b_i  +  b_j^\dagger 
b_j + b_i^\dagger b_j^\dagger b_j  b_i \right),
\label{u1_mf_holon_result}
\eqa
with $\Delta^f_{ij} = f_{j\uparrow}f_{i\downarrow}-f_{j\downarrow} 
f_{i\uparrow}$, the spinon pairing field.
The third term in Eq.(\ref{u1_tjmodel0}) represents the intersite spin (spinon) 
interaction and is rewritten,
\bqa
&& -\frac{J}{2} \Bigl< b_i b_j b_j^{\dagger}b_i^{\dagger}\Bigr> 
(f_{i\downarrow}^{\dagger}f_{j\uparrow}^{\dagger}-f_{i\uparrow}^ 
{\dagger}f_{j\downarrow}^{\dagger})(f_{j\uparrow}f_{i\downarrow}-f_{j\downarrow} 
f_{i\uparrow}) \nn
& = & -\frac{J_p}{2} 
(f_{i\downarrow}^{\dagger}f_{j\uparrow}^{\dagger}-f_{i\uparrow}^ 
{\dagger}f_{j\downarrow}^{\dagger})(f_{j\uparrow}f_{i\downarrow}-f_{j\downarrow} 
f_{i\uparrow}),
\label{u1_mf_spinon_result}
\eqa
where $J_p = J(  1 + <b_i^\dagger b_i> + <b_j^\dagger b_j> + <b_i^\dagger b_i 
b_j^\dagger b_j >  )$ or $J_p = J(1-x)^2$ with $x$, 
the uniform 
hole doping concentration.
The fourth term in Eq.(\ref{u1_tjmodel0}) is written,
\bqa
\lefteqn{ \frac{J}{2} \Bigl< b_i b_j b_j^{\dagger}b_i^{\dagger} \Bigr> \Bigl< 
(f_{i\downarrow}^{\dagger}f_{j\uparrow}^{\dagger}-f_{i\uparrow}^{\dagger}
f_{j\downarrow}^{\dagger})(f_{j\uparrow}f_{i\downarrow}-f_{ j\downarrow} 
f_{i\uparrow}) \Bigr> } \nn
& = & \frac{J}{2}  \left( 1 + <b_i^\dagger b_i> + <b_j^\dagger b_j> + 
<b_i^\dagger b_i b_j^\dagger b_j > \right) 
<|\Delta^f_{ij}|^2>. \nn
\label{u1_mf_const}
\eqa

The intersite spinon interaction term in Eq.(\ref{u1_mf_spinon_result}) is 
decomposed into the direct, exchange and pairing 
channels\cite{UBBENS},
\bqa
&& -\frac{J_p}{2} 
(f_{i\downarrow}^{\dagger}f_{j\uparrow}^{\dagger}-f_{i\uparrow}^ 
{\dagger}f_{j\downarrow}^{\dagger})(f_{j\uparrow}f_{i\downarrow}-f_{j\downarrow} 
f_{i\uparrow}) \nn
& = & \frac{J_p}{4} \Bigl[ \sum_{k=1}^3 (f_{i\alpha}^{\dagger} \sigma_{\alpha 
\beta} ^k f_{i\beta})( f_{j\gamma }^{\dagger} 
\sigma_{\gamma 
\delta} ^k f_{j\delta} ) - ( f_{i\alpha}^\dagger \sigma^0_{\alpha \beta} 
f_{i\beta} ) ( f_{i\gamma}^\dagger \sigma^0_{\gamma \delta} 
f_{j\delta} 
) \Bigr] \nn
& = & v_D + v_E + v_P 
\label{eq:mf_1} 
\eqa
with $\sigma^0 = I$, the identity matrix and $\sigma^{1,2,3}$, the Pauli spin 
matrices,
where $v_D$, $v_E$ and $v_P$ are the spinon interaction terms of the direct, 
exchange and pairing channels respectively\cite{LEE2},
\bqa
v_D & = & -\frac{J_p}{8} \sum_{k=0}^{3} ( f_i^{\dagger} \sigma^k f_i ) 
( f_j^{\dagger} \sigma^k f_j ),  \label{vD} \\
v_E & =  & -\frac{J_p}{4} \Bigl( 
(f_{i\sigma}^{\dagger}f_{j\sigma})(f_{j\sigma}^{\dagger}f_{i\sigma}) - n_i 
\Bigr), \label{vE} \\
v_P & = &  -\frac{J_p}{2} 
(f_{i\downarrow}^{\dagger}f_{j\uparrow}^{\dagger}
-f_{i\uparrow}^{\dagger}f_{j\downarrow}^{\dagger}) 
(f_{j\uparrow}f_{i\downarrow}-f_{j\downarrow}f_{i\uparrow}). \label{vP}
\eqa
Here $\sigma^0$ is the unit matrix and $\sigma^{1,2,3}$, the Pauli spin 
matrices.

Combining Eq.(\ref{u1_mf_holon_result}) and Eq.(\ref{u1_mf_const}), we have
\bqa
&& -\frac{J}{2}<|\Delta_{ij}^f|^2> \left( 1 + b_i^\dagger b_i  +  b_j^\dagger 
b_j + b_i^\dagger b_j^\dagger b_j  b_i \right) \nn
&& + \frac{J}{2} <|\Delta_{ij}^f|^2> \left( 1 + <b_i^\dagger b_i> + <b_j^\dagger 
b_j> + <b_i^\dagger b_i b_j^\dagger b_j > \right) 
\nn
&& = -\frac{J}{2} <|\Delta_{ij}^f|^2>  b_i^\dagger b_j^\dagger b_j  b_i
+ \frac{J}{2} <|\Delta^f_{ij}|^2> <b_i^\dagger b_i b_j^\dagger b_j >  \nn
&& -\frac{J}{2} <|\Delta_{ij}^f|^2> \left[ \left(b_i^\dagger b_i -<b_i^\dagger 
b_i>\right) + \left(b_j^\dagger b_j-<b_j^\dagger b_j> 
\right) 
\right].
\label{u1_mf_holon_const}
\eqa

Collecting the decomposed terms Eq.(\ref{u1_mf_holon_result}) through 
Eq.(\ref{u1_mf_const}) in association with Eqs.(\ref{eq:mf_1}) 
through 
Eq.(\ref{u1_mf_holon_const}), we write 
\bqa
&&H_J= 
-\frac{J}{2} \sum_{<i,j>} |\Delta_{ij}^f|^2  b_i^\dagger b_j^\dagger b_j  b_i 
\nn
&& -J_p \sum_{<i,j>} \Bigl[ 
\frac{1}{2} 
(f_{i\downarrow}^{\dagger}f_{j\uparrow}^{\dagger}
-f_{i\uparrow}^{\dagger}f_{j\downarrow}^{\dagger}) 
(f_{j\uparrow}f_{i\downarrow}-f_{j\downarrow}f_{i\uparrow})  \nn
&& +\frac{1}{4} \Bigl( 
(f_{i\sigma}^{\dagger}f_{j\sigma})(f_{j\sigma}^{\dagger}f_{i\sigma}) - n_i 
\Bigr) \nn
&& + \frac{1}{8} \sum_{k=0}^{3} ( f_i^{\dagger} \sigma^k f_i ) ( f_j^{\dagger} 
\sigma^k f_j ) 
\Bigr] \nn
&& + \frac{J}{2} \sum_{<i,j>} |\Delta^f_{ij}|^2 <b_i^\dagger b_i> <b_j^\dagger 
b_j >  \nn
&& -\frac{J}{2} \sum_{<i,j>} |\Delta_{ij}^f|^2 \left[ \left(b_i^\dagger b_i 
-<b_i^\dagger b_i>\right) + \left(b_j^\dagger 
b_j-<b_j^\dagger b_j> 
\right) \right],
\label{u1_tjmodel}
\end{eqnarray}
where we considered $<|\Delta_{ij}^f|^2> = |\Delta_{ij}^f|^2$ and ignored the 
fifth term in Eq.(\ref{u1_tjmodel0}).

As are shown in Eqs.(\ref{vD}) through (\ref{vP}) the spinon interaction term is 
decomposed into the direct, exchange and pairing 
channels 
respectively.
Proper Hubbard-Stratonovich transformations corresponding to these channels and 
saddle point approximation leads to the effective 
Hamiltonian,
\bqa
H_{eff} & = & 
\frac{J_p}{4} \sum_{<i,j>} \Bigl[ |\chi_{ij}|^2 - \chi_{ij}^* ( 
f_{i\sigma}^{\dagger}f_{j\sigma} + \frac{4t}{J_p} b_i^\dagger b_j )  
- c.c. 
\Bigr]  \nn
&& - \frac{J}{2} \sum_{<i,j>} |\Delta^f_{ij}|^2 b_{j}^* b_{i}^* b_{j}b_{i} \nn
&& + \frac{J_p}{2} \sum_{<i,j>} \Bigl[ |\Delta^f_{ij}|^2 - \Delta^{f}_{ij} 
(f_{i\downarrow}^\dagger f_{j\uparrow}^\dagger - 
f_{i\uparrow}^\dagger  f_{j\downarrow}^\dagger )  - c.c. \Bigr] \nn
&& + \frac{J_p}{2} \sum_{<i,j>} \sum_{l=0}^{3} \Bigl[ (\rho^l_{j})^2 -  
\rho^l_{j} ( f_i^{\dagger} \sigma^l f_i )  \Bigr]  + 
\frac{J_p}{2} 
\sum_{i} (f_{i\sigma}^\dagger f_{i\sigma}) \nn
&& + \frac{4t^2}{J_p} \sum_{<i,j>} ( b_i^\dagger b_j )(b_j^\dagger b_i ) \nn
&& + \frac{J}{2} \sum_{<i,j>} |\Delta^f_{ij}|^2 <b_i^\dagger b_i> <b_j^\dagger 
b_j >  \nn
&&  -\frac{J}{2} \sum_{<i,j>} |\Delta_{ij}^f|^2 \left[ \left(b_i^\dagger b_i - 
<b_i^\dagger b_i>\right) + \left(b_j^\dagger b_j - 
<b_j^\dagger 
b_j> \right) \right] \nn
&&  - \mu \sum_{i} f_{i\sigma}^{\dagger}f_{i\sigma}( 1+b_i^\dagger b_i) - 
i\sum_{i} \lambda_{i}(f_{i\sigma}^{\dagger}f_{i\sigma} + 
b_{i}^{\dagger}b_{i} -1),
\label{u1_eff_hamil1}
\eqa
where $\Delta_{ij}^{b} = <b_{i}b_{j}>$, $\chi_{ij}= < 
f_{i\sigma}^{\dagger}f_{j\sigma} + \frac{4t}{J_p} b_{i}^{\dagger}b_{j}>$, 
$\Delta^f_{ij} = 
< f_{j\uparrow}f_{i\downarrow}-f_{j\downarrow}f_{i\uparrow} >$ and 
$\rho_{i}^{k}=<\frac{1}{2}f_{i}^\dagger \sigma^k f_i>$ are proper 
saddle 
points.

We note that $\rho^l_{i} = \frac{1}{2} < f_i^{\dagger} \sigma^l f_i > = < S^l_i 
> = 0 $ for $l=1,2,3$, $\rho^0_{i} = \frac{1}{2} < 
f_{i 
\sigma}^\dagger f_{i \sigma} > = \frac{1}{2}$ for $l=0$ for the contribution of 
the direct spinon interaction term (the fourth 
term). 
The expression $(b_{j}^{\dagger}b_{i})(b_{i}^{\dagger}b_{j})$ in the fifth term 
of Eq.(\ref{u1_eff_hamil1}) represents the exchange 
interaction 
channel. 
The exchange channel will be ignored owing to a large cost in energy, $U \approx 
\frac{4t^2}{J}$\cite{UBBENS}\cite{WEN}.
The resulting effective Hamiltonian is
\bq
H^{MF} = H^{\Delta,\chi} + H^b + H^f,
\label{mean_H_u1}
\eq
where $H^{\Delta,\chi}$ represents the the saddle point energy involved with the 
spinon pairing order parameter $\Delta^f$ and the 
hopping order 
parameter $\chi$,
\bqa
H^{\Delta,\chi} & = & 
 J \sum_{<i,j>} \frac{1}{2} |\Delta^f_{ij}|^2 x^2 + \frac{J_p}{2} \sum_{<i,j>} 
\Bigl[ |\Delta_{ij}^{f}|^{2} + \frac{1}{2} 
|\chi_{ij}|^{2} + 
\frac{1}{4} \Bigr], 
\label{u1_Delta_chi}
\eqa
$H^b$ is the holon Hamiltonian,
\bqa
H^b  & = & -t \sum_{<i,j>} \Bigl[ \chi_{ij}^{*}(b_{i}^{\dagger}b_{j}) + c.c.  
\Bigr] \nn
     & - & \sum_{<i,j>} \frac{J}{2}|\Delta^f_{ij}|^2 b_{i}^\dagger b_{j}^\dagger 
 b_j b_i \nn
    & - & \sum_{i} \mu^b_i ( b_{i}^{\dagger}b_{i} -x ),
\label{u1_holon_sector}
\eqa
where $\mu^b_i = i\lambda_i + \frac{J}{2}\sum_{j=i\pm \hat x, i \pm \hat y} 
|\Delta_{ij}^f|^2$
and $H^f$, the spinon Hamiltonian,
\bqa
H^f & = & 
-  \frac{J_p}{4} \sum_{<i,j>} \Bigl[ \chi_{ij}^{*} 
(f_{i\sigma}^{\dagger}f_{j\sigma}) + c.c. \Bigr] \nn
& - & \frac{J_p}{2} \sum_{<i,j>} \Bigl[ \Delta_{ij}^{f*} 
(f_{j\uparrow}f_{i\downarrow}-f_{j\downarrow}f_{i\uparrow}) + c.c. \Bigr]   
\nn
& - & \sum_{i} \mu^f_i \left( f_{i\sigma}^{\dagger} f_{i\sigma} -(1-x) \right),
\label{u1_spinon_sector}
\eqa
where $\mu^f_i = \mu (1-x)  + i\lambda_i$.

\renewcommand{\theequation}{B\arabic{equation}}
\setcounter{equation}{0}
\section*{APPENDIX B: Derivation of the SU(2) holon Hamiltonian} 
The SU(2) slave-boson representation of the above t-J Hamiltonian 
reads\cite{LEE2}
\bqa
H  & =  &  - \frac{t}{2} \sum_{<i,j>\sigma}  \Bigl[ (f_{\sigma 
i}^{\dagger}f_{\sigma j})(b_{1j}^{\dagger}b_{1i}-b_{2i}^{\dagger}b_{2j}) + c.c.  
\nn
&& + (f_{2i}f_{1j}-f_{1i}f_{2j}) (b_{1j}^{\dagger}b_{2i} + 
b_{1i}^{\dagger}b_{2j}) + c.c. \Bigr] \nn
 && -  \frac{J}{2} \sum_{<i,j>} ( 1 + h_{i}^\dagger h_{i} ) ( 1 + h_{j}^\dagger 
h_{j} ) \times \nn
  && (f_{2i}^{\dagger}f_{1j}^{\dagger}-f_{1i}^ 
{\dagger}f_{2j}^{\dagger})(f_{1j}f_{2i}-f_{2j} f_{1i}) -  \mu_0 \sum_i  
h_i^\dagger 
h_i  \nn
 && -  \sum_i  \Bigl[ i\lambda_{i}^{(1)} ( f_{1i}^{\dagger}f_{2i}^{\dagger} + 
b_{1i}^{\dagger}b_{2i}) + i \lambda_{i}^{(2)} ( 
f_{2i}f_{1i} + 
b_{2i}^\dagger b_{1i} ) \nn
  && + i \lambda_{i}^{(3)} ( f_{1i}^{\dagger}f_{1i} -  f_{2i} f_{2i}^{\dagger} + 
b_{1i}^{\dagger}b_{1i} - b_{2i}^{\dagger}b_{2i} ) 
\Bigr],
\label{eq:su2_sb_representation}
\eqa
where $\lambda_{i}^{(1),(2),(3)}$ are the real Lagrangian multipliers to enforce 
the local single occupancy constraint in the SU(2) 
slave-boson 
representation\cite{WEN}.
Taking decomposition of the Heisenberg interaction term above into terms 
involving charge and spin fluctuations separately, 
uncorrelated mean 
field contributions and correlated fluctuations, i.e., correlations between 
charge and spin fluctuations as in the U(1) case, the 
SU(2) 
Hamiltonian is rewritten,
\bqa
H^{SU(2)}_{t-J} &=& - \frac{t}{2} \sum_{<i,j>}  \Bigl[ (f_{i 
\alpha}^{\dagger}f_{j 
\alpha})(b_{j1}^{\dagger}b_{i1}-b_{i2}^{\dagger}b_{j2}) + 
c.c. \nn 
&& + (f_{i2}f_{j1}-f_{i1}f_{j2}) (b_{j1}^{\dagger}b_{i2} + 
b_{i1}^{\dagger}b_{j2}) + c.c. \Bigr] \nn 
&& -\frac{J}{2} \sum_{<i,j>} \Bigl[
\Bigl< (f_{i2}^{\dagger}f_{j1}^{\dagger}-f_{i1}^ 
{\dagger}f_{j2}^{\dagger})(f_{j1}f_{i2}-f_{j2} f_{i1}) \Bigr> ( 1 + h_i^\dagger 
h_i) ( 1 +  
h_j^{\dagger}h_j ) \nn
&& + \Bigl< ( 1 + h_i^\dagger h_i) ( 1 +  h_j^{\dagger}h_j ) \Bigr> 
(f_{i2}^{\dagger}f_{j1}^{\dagger}-f_{i1}^ 
{\dagger}f_{j2}^{\dagger})(f_{j1}f_{i2}-f_{j2} f_{i1})  \nn
 & & -  \Bigl< ( 1 + h_i^\dagger h_i) ( 1 +  h_j^{\dagger}h_j ) \Bigr> \Bigl< 
(f_{i2}^{\dagger}f_{j1}^{\dagger}-f_{i1}^ 
{\dagger}f_{j2}^{\dagger})(f_{j1}f_{i2}-f_{j2} f_{i1}) \Bigr> \nn
  & & + \Bigl( ( 1 + h_i^\dagger h_i) ( 1 +  h_j^{\dagger}h_j ) - \Bigl< ( 1 + 
h_i^\dagger h_i) ( 1 +  h_j^{\dagger}h_j ) \Bigr> 
\Bigr) \times 
\nn
  && \Bigl( (f_{i2}^{\dagger}f_{j1}^{\dagger}-f_{i1}^ {\dagger}f_{j2}^{\dagger} 
) ( f_{j1}f_{i2}-f_{j2} f_{i1}) - 
\Bigl<(f_{i2}^{\dagger}f_{j1}^{\dagger}-f_{i1}^ 
{\dagger}f_{j2}^{\dagger})(f_{j1}f_{i2}-f_{j2} f_{i1})\Bigr> \Bigr) \Bigr] \nn
&& - \mu \sum_i ( 1 - h_i^{\dagger} h_i ) \nn
&& -  \sum_i  \Bigl(
 i\lambda_{i}^{(1)} ( f_{i1}^{\dagger}f_{i2}^{\dagger} + b_{i1}^{\dagger}b_{i2}) 
+ i \lambda_{i}^{(2)} ( f_{i2}f_{i1} + b_{i2}^\dagger b_{i1} ) \nn
 && + i \lambda_{i}^{(3)} ( f_{i1}^{\dagger}f_{i1} -  f_{i2} f_{i2}^{\dagger} + 
b_{i1}^{\dagger}b_{i1} - b_{i2}^{\dagger}b_{i2} ) 
\Bigr).
\label{su2_tjmodel2}
\eqa
The intersite holon interaction (the third term in Eq.(\ref{su2_tjmodel2})) is 
rewritten,
\bqa
&& -\frac{J}{2} \Bigl< (f_{i2}^{\dagger}f_{j1}^{\dagger}-f_{i1}^ 
{\dagger}f_{j2}^{\dagger})(f_{j1}f_{i2}-f_{j2} f_{i1}) \Bigr> ( 1 + 
h_i^\dagger 
h_i) ( 1 +  h_j^{\dagger}h_j ) \nn
&& =  -\frac{J}{2} <|\Delta^f_{ij}|^2> ( 1 + h_i^\dagger h_i +  h_j^{\dagger}h_j 
+ h_i^\dagger h_i h_j^{\dagger}h_j ) \nn
&& \approx  -\frac{J}{2} |\Delta^f_{ij}|^2 ( 1 + b_{i\alpha}^\dagger b_{i\alpha} 
+  b_{j\alpha}^\dagger b_{j\alpha}  + 
b_{i\alpha}^{\dagger}b_{j\beta}^{\dagger}b_{j\beta}b_{i\alpha}  ) \nn
\label{eq:su2_mf_holon_result}
\eqa
where $\Delta^f_{ij} = \Bigl<f_{j1}f_{i2}-f_{j2} f_{i1} \Bigr>$ is the spinon 
singlet pairing order parameter.
The intersite spinon interaction (the fourth term in Eq.(\ref{su2_tjmodel2})) is 
rewritten in terms of decomposed 
Hatree-Fock-Bogoliubov 
channels in the same way as in the U(1) case, 
\bq
-\frac{J_p}{2} (f_{i2}^{\dagger} f_{j1}^{\dagger}-f_{i1}^ 
{\dagger}f_{j2}^{\dagger})(f_{j1}f_{i2}-f_{j2} f_{i1})  = v_D + v_E + v_P,
\label{eq:mf_spinon_spinon} 
\eq
where $v_D$, $v_E$ and  $v_P$ are the interactions corresponding to the direct, 
exchange and pairing channels respectively,
\bqa
v_D & = & -\frac{J_p}{8} \sum_{l=0}^{3} ( f_i^{\dagger} \sigma^l f_i ) ( 
f_j^{\dagger} \sigma f_j ), \label{su2_vD} \\
 v_E & = & -\frac{J_p}{4} \Bigl( 
(f_{i\sigma}^{\dagger}f_{j\sigma})(f_{j\sigma}^{\dagger}f_{i\sigma}) - n_i 
\Bigr), \label{su2_vE} 
\\
 v_P & = & -\frac{J_p}{2} 
(f_{i2}^{\dagger}f_{j1}^{\dagger}-f_{i1}^{\dagger}f_{j2}^{\dagger}) 
(f_{j1}f_{i2}-f_{j2}f_{i1}). 
\label{su2_vP}
\eqa
Here $\sigma^0$ is the unit matrix and $\sigma^{1,2,3}$, the Pauli spin 
matrices.
The fifth term in Eq.(\ref{su2_tjmodel2}) is written,
\bqa
&& \frac{J}{2} \Bigl<  ( 1 + h_i^\dagger h_i) ( 1 +  h_j^{\dagger}h_j ) \Bigr> 
\times \nn
&& \Bigl< (f_{i2}^{\dagger}f_{j1}^{\dagger}-f_{i1}^ {\dagger}f_{j2}^{\dagger}) 
(f_{j1}f_{i2}-f_{j2} f_{i1}) \Bigr> \nn
& \approx & \frac{J}{2} \Bigl<  ( 1 + h_i^\dagger h_i) \Bigr> \Bigl< ( 1 +  
h_j^{\dagger}h_j) \Bigr> \times \nn
&& \Bigl< (f_{i2}^{\dagger}f_{j1}^{\dagger}-f_{i1}^ {\dagger}f_{j2}^{\dagger}) 
\Bigr> \Bigl< (f_{j1}f_{i2}-f_{j2} f_{i1}) \Bigr>  
\nn
& = & \frac{J}{2} |\Delta^f_{ij}|^2 ( 1 + < h_i^\dagger h_i > + < h_j^{\dagger} 
h_j > + < h_i^\dagger h_i > < h_j^{\dagger}h_j > ). 
\nn
\label{eq:su2_mf_const}
\eqa
We introduced $\Bigl< (f_{i2}^{\dagger}f_{j1}^{\dagger}-f_{i1}^ 
{\dagger}f_{j2}^{\dagger})(f_{j1}f_{i2}-f_{j2} f_{i1}) \Bigr> 
\approx \Bigl< 
(f_{i2}^{\dagger}f_{j1}^{\dagger}-f_{i1}^ {\dagger}f_{j2}^{\dagger}) \Bigr> 
\Bigl< (f_{j1}f_{i2}-f_{j2} f_{i1}) \Bigr> = 
|\Delta^f_{ij}|^2$ and 
$ \Bigl< ( 1 + h_i^\dagger h_i) ( 1 +  h_j^{\dagger}h_j ) \Bigr> \approx \Bigl< 
( 1 + h_i^\dagger h_i)\Bigr> \Bigl< ( 1 +  
h_j^{\dagger}h_j ) 
\Bigr>$.

By introducing the Hubbard-Stratonovich fields, 
${\rho}_{i}^{k}$, $\chi_{ij}$ and $\Delta_{ij}$ for the spinon direct, exchange 
and pairing order shown in Eqs.(\ref{su2_vD}), 
(\ref{su2_vE}), 
(\ref{su2_vP}),
we rewrite the effective Hamiltonian for Eq.(\ref{su2_tjmodel2}), 
\begin{eqnarray}
&& H_{SU(2)}^{MF}  = 
\frac{J_p}{4} \sum_{<i,j>} \Bigl[ |\chi_{ij}|^2 - \chi_{ij}^* \{ f_{i 
\sigma}^{\dagger}f_{ \sigma j}  + \frac{2t}{J_p} 
(b_{i1}^{\dagger}b_{j1}-b_{j2}^{\dagger}b_{2i }) \}  - c.c. \Bigr] \nn
& - &  \frac{J}{2} \sum_{<i,j>} |\Delta^f_{ij}|^2 \Bigl[ b_{i\alpha}^\dagger 
b_{j\beta}^\dagger b_{j\alpha} b_{i\beta}  \Bigr] \nn
 & + & \frac{J_p}{2} \sum_{<i,j>} \Bigl[ |\Delta_{ij}|^2 - \Delta_{ij} \{ 
(f_{i2}^{\dagger}f_{j1}^{\dagger}-f_{i1}^{\dagger}f_{j2}^{\dagger}) - 
\frac{t}{J_p} (b_{j1}^{ \dagger}b_{i2} + b_{i1}^{\dagger}b_{j2}) \}  - c.c. 
\Bigr]  \nn
&+&  \frac{J_p}{2} \sum_{<i,j>} \sum_{l=0}^{3} \Bigl( (\rho^l_{ij})^2 - 
\rho^l_{ij} ( f_i^{\dagger} \sigma^l f_i ) \Bigr) \nn
& + &  \frac{t^2}{J_p}  \sum_{<i,j>} \Bigl[ 
(b_{i1}^{\dagger}b_{j1}-b_{j2}^{\dagger}b_{i2}) 
(b_{j1}^{\dagger}b_{i1}-b_{i2}^{\dagger}b_{j2})  \nn
&& + \frac{1}{2} (b_{j1}^{\dagger}b_{i2}+b_{i1}^{\dagger}b_{j2}) 
(b_{i2}^{\dagger}b_{j1} + b_{j2}^{\dagger}b_{i1}) \Bigr]  \nn
& - & \frac{J}{2} \sum_{<i,j>} |\Delta^f_{ij}|^2 \Bigl[  h_j^{\dagger} h_j + 
h_{i}^\dagger h_{i} - <h_j^{\dagger} h_j > - 
<h_{i}^\dagger h_{i} > 
  \Bigr]  + \frac{J}{2} \sum_{<i,j>} |\Delta^f_{ij}|^2 x^2 \nn
& + & \frac{J_p}{2} \sum_{i} (f_{i \sigma}^\dagger f_{i \sigma})
- \mu \sum_i ( 1 - h_i^{\dagger} h_i ) \nn
& - & \sum_i  \Bigl[
 i \lambda_{i}^{1} ( f_{i1}^{\dagger}f_{i2}^{\dagger} + b_{i1}^{\dagger}b_{i2})
+ i \lambda_{i}^{2} ( f_{i2}f_{i1} + b_{i2}^\dagger b_{i1} ) \nn
&& + i \lambda_{i}^{3} ( f_{i1}^{\dagger}f_{i1} -  f_{i2} f_{i2}^{\dagger} + 
b_{i1}^{\dagger}b_{i1} - b_{i2}^{\dagger}b_{i2} )], 
\label{su2_mf_hamiltonian1}
\end{eqnarray}
where 
$\chi_{ij} = < f_{i\sigma}^{\dagger}f_{j\sigma} + \frac{2t}{J_p} 
(b_{i1}^{\dagger}b_{j1}-b_{j2}^{\dagger}b_{i2}) >$,
$\Delta_{ij} = \Bigl< (f_{i1}f_{j2} - f_{i2}f_{j1}) - \frac{t}{J_p} ( 
b_{i2}^\dagger b_{j1} + b_{j2}^\dagger b_{i1} ) \Bigr> = 
\Delta_{ij}^f - 
\frac{t}{J(1-x)} \chi_{ij;12}^b$, with $\chi_{ij;12}^b = \Bigl< b_{i2}^\dagger 
b_{j1} + b_{j2}^\dagger b_{i1} \Bigr>$ and $x$, the 
hole doping 
rate.

To simplify the effective Hamiltonian, we rearrange each term in 
Eq.(\ref{su2_mf_hamiltonian1}).
For the paramagnetic state, we obtain $\rho^l_{i} = \frac{1}{2} ( f_i^{\dagger} 
\sigma^l f_i ) = < S^l_i > = 0 $ for $l=1,2,3$ and 
$\rho^0_{i} = 
\frac{1}{2} < f_{i \sigma}^\dagger f_{i \sigma} > = \frac{1}{2}$  for $l=0$. 
The one-body holon term (the sixth term) in the above Hamiltonian is 
incorporated into the effective chemical potential term.
By setting $\Delta_{ij} = \Delta_{ij}^f - \frac{t}{J(1-x)} \chi_{ij;12}^b$ where 
$\Delta_{ij}^f = < f_{i1}f_{j2} - f_{i2}f_{j1}>$ 
and 
$\chi_{ij;12}^b = \Bigl< b_{i2}^\dagger b_{j1} + b_{j2}^\dagger b_{i1} \Bigr>$,
we rearrange the third term and the second term in the bracket of the fifth term 
to obtain the effective Hamiltonian,
\begin{eqnarray}
&& H^{MF}_{SU(2)} = 
\frac{J}{2} \sum_{<i,j>} |\Delta^f_{ij}|^2 x^2   +\frac{J_p}{2} \sum_{<i,j>} 
\Bigl[ |\Delta_{ij}^{f}|^{2} + \frac{1}{2} 
|\chi_{ij}|^{2} + 
\frac{1}{4} \Bigr] \nn
& & -\frac{t}{2} \sum_{<i,j>} \Bigl[ \chi_{ij}^* (b_{i1}^{\dagger}b_{j1} - 
b_{j2}^{\dagger}b_{i2}) -\Delta^f_{ij} 
(b_{j1}^{\dagger}b_{i2} + 
b_{i1}^{\dagger}b_{j2})\Bigr] - c.c. \nn
&& -\frac{J_p}{4} \sum_{<i,j>} \Bigl[ \chi_{ij}^* (f_{i \sigma}^{\dagger}f_{j 
\sigma}) + c.c. \Bigr]  \nn
& - &  \frac{J}{2} \sum_{<i,j>} |\Delta^f_{ij}|^2 \Bigl[ b_{i\alpha}^\dagger 
b_{j\beta}^\dagger b_{j\alpha} b_{i\beta}  \Bigr] \nn
& & -\frac{J_p}{2} \sum_{<i,j>} \Bigl[ \Delta_{ij}^{f*} 
(f_{j1}f_{i2}-f_{j2}f_{i1}) + c.c. \Bigr] \nn
&& -\sum_{i} \mu^b_{i} ( h_{i}^{\dagger}h_{i} - x ) \nn
&& - \sum_i \Bigl[ i\lambda_{i}^{1} ( f_{i1}^{\dagger}f_{i2}^{\dagger} + 
b_{i1}^{\dagger}b_{i2}^{\dagger}) + i \lambda_{i}^{2} ( 
f_{i2}f_{i1} + 
b_{i2}b_{i1} ) \nn
&& + i \lambda_{i}^{3} ( f_{i1}^{\dagger}f_{i1} -  f_{i2} f_{i2}^{\dagger} + 
b_{i1}^{\dagger}b_{i1} - b_{i2}^{\dagger}b_{i2} ) 
\Bigr] \nn
&& - \frac{t}{2} \sum_{<i,j>}  \left( \Delta^{f}_{ij} - (f_{j1}f_{i2}- 
f_{j2}f_{i1}) \right) \chi_{ij;12}^{b*} - c.c. \nn
&& + \frac{t^2}{2J_p} \sum_{<i,j>}  \left| \chi_{ij;12}^b - (b_{i2}^\dagger 
b_{j1} + b_{j2}^\dagger b_{i1} ) \right|^2  \nn
&& + \frac{t^2}{J_p}\sum_{<i,j>}  (b_{i1}^{\dagger}b_{j1} - 
b_{j2}^{\dagger}b_{i2}) ( b_{j1}^{\dagger}b_{i1} - 
b_{i2}^{\dagger}b_{j2}),
\label{su2_mf_hamiltonian3}
\end{eqnarray}
where $\mu^b_i = -\mu + \frac{J}{2} \sum_{j=i\pm \hat x, i \pm \hat y} 
|\Delta^f_{ij}|^2$.

We neglect correlations between the fluctuations of order parameters.
This is because correlations between the spin (spinon pair) and charge (holon 
pair) fluctuations are expected to be small as 
compared to the 
saddle point contribution of the order parameters (the first  and second terms 
in the above Hamiltonian) particularly near the 
pseudogap 
temperature and the bose condensation temperature.
However, individual fluctuations of the spinon pairing and holon pairing order 
parameters are not ignored.
We also neglect the fluctuations of order parameters (ninth and tenth terms) in 
Eq.(\ref{su2_mf_hamiltonian3}).
The ninth term represents the fluctuation of spinon pairing order parameter and 
we neglect it owing to its vanishment as an 
expectation value.
The tenth term represents the fluctuations of holon hopping order parameter and 
is negligible at low temperature as its fluctuations 
die out in 
the low temperature regions where pairing order parameters $\Delta^f$ begins to 
open.
Owing to the high energy cost involved with the Coulomb repulsion energy the 
exchange interaction terms (the last positive energy 
terms) will be 
ignored\cite{UBBENS}-\cite{WEN}.
We then obtain the mean field Hamiltonian, $H = H^{\Delta,\chi} + H^b + H^f$,
where $H^{\Delta,\chi}$ is the saddle point contribution of order parameters, 
$\chi$ and $\Delta^f$,
\bqa
H^{\Delta,\chi}_{SU(2)} & = &  
\frac{J}{2} \sum_{<i,j>} |\Delta^f_{ij}|^2 x^2  + \frac{J_p}{2} \sum_{<i,j>} 
\Bigl[ |\Delta_{ij}^{f}|^{2} + \frac{1}{2} |\chi_{ij} 
|^{2} + 
\frac{1}{4} \Bigr],
\label{su2_mean}
\eqa
$H^b$ is the holon Hamiltonian,
\bqa
&& H^b_{SU(2)}  =   
-\frac{t}{2} \sum_{<i,j>} \Bigl[ \chi_{ij}^* (b_{i1}^{\dagger}b_{j1} - 
b_{j2}^{\dagger}b_{i2}) \nn
&& -\Delta^f_{ij} (b_{j1}^{\dagger}b_{i2} + b_{i1}^{\dagger}b_{j2})\Bigr] - c.c. 
\nn
& - &  \frac{J}{2} \sum_{<i,j>} |\Delta^f_{ij}|^2 \Bigl[ b_{i\alpha}^\dagger 
b_{j\beta}^\dagger b_{j\alpha} b_{i\beta}  \Bigr] \nn
&& -\sum_{i} \Bigl[ \mu^b_{i} ( h_{i}^{\dagger}h_{i} - x ) + i\lambda_{i}^{1} ( 
b_{i1}^{\dagger}b_{i2}^{\dagger})  + i 
\lambda_{i}^{2} ( 
b_{i2}b_{i1} ) \nn
&& + i \lambda_{i}^{3} ( b_{i1}^{\dagger}b_{i1} - b_{i2}^{\dagger}b_{i2} ) 
\Bigr],
\label{su2_holon_sector}
\eqa
and $H^f$, the spinon Hamiltonian,
\bqa
&& H^f_{SU(2)}  =  
-  \frac{J_p}{4} \sum_{<i,j>} \Bigl[ \chi_{ij}^* (f_{i \sigma}^{\dagger}f_{j 
\sigma}) + c.c. \Bigr]  \nn
&& -\frac{J_p}{2} \sum_{<i,j>} \Bigl[ \Delta_{ij}^{f*} 
(f_{j1}f_{i2}-f_{j2}f_{i1}) + c.c. \Bigr], \nn 
&& - \sum_i \Bigl[ i\lambda_{i}^{1} ( f_{i1}^{\dagger}f_{i2}^{\dagger} )  + i 
\lambda_{i}^{2} ( f_{i2}f_{i1} ) + i \lambda_{i}^{3} ( 
f_{i1}^{\dagger}f_{i1} - f_{i2} f_{i2}^{\dagger} ) \Bigr], \nn
\label{su2_spinon_sector}
\eqa
where 
$\chi_{ij}= < f_{i \sigma}^{\dagger}f_{j \sigma} + \frac{2t}{J_p} 
(b_{i1}^{\dagger}b_{j1} - b_{j2}^\dagger b_{i2} )>$,
$\Delta_{ij}^{f}=< f_{j1}f_{i2}-f_{j2}f_{i1} >$ and
$\mu^b_i = -\mu - \frac{J}{2} \sum_{j=i\pm \hat x, i \pm \hat y} 
|\Delta^f_{ij}|^2$.
Taking the saddle point value for the Lagrangian multiplier fields 
$\lambda_{i}^{l} = 0$\cite{LEE2}, we obtain the holon 
Hamiltonian, 
Eq.(\ref{eq:su2_model}).

\renewcommand{\theequation}{C\arabic{equation}}
\setcounter{equation}{0}
\section*{APPENDIX C: Derivation of Eqs.(4) and (5)} 
Introducing a uniform hopping order parameter, $\chi_{ji}=\chi$, and a $d$-wave 
spinon pairing order parameters, $ \Delta_{ji}^{f}= 
\pm 
\Delta_f$(the sign $+(-)$ is for the ${\bf ij}$ link parallel to $\hat x$ ($\hat 
y$)), 
we obtain the energy-momentum space representation of the action for the U(1) 
theory,
\bqa
&& S^b_{U(1)}  =  \sum_{{\bf k},\omega_n} 
( -i \omega_n + \epsilon( {\bf k} ) ) b({\bf k},\omega_n)^\dagger b({\bf k}, 
\omega_n )\nn
&& + \frac{1}{2N \beta} \sum_{{\bf k}, {\bf k}^{'}, {\bf q}} 
\sum_{\omega_n, \omega_n^{'}, \nu_n}
v( {\bf k} - {\bf k}^{'})
b(-{\bf k}^{'} + {\bf q},-\omega_n^{'}+\nu_n)^\dagger b({\bf 
k}^{'},\omega_n^{'})^\dagger b({\bf k},\omega_n) b(-{\bf k} + {\bf q}, 
-\omega_n+\nu_n),
\label{eq:u1_model_momentum}
\eqa
where 
$ b({\bf k},\omega_n) = \frac{1}{\sqrt{N\beta}} \int_0^\beta d \tau \sum_{i} 
e^{i( \omega_n \tau  - {\bf k} \cdot {\bf r}_i)} 
b_i(\tau)$,
$\epsilon( {\bf k} ) = -2t\chi \gamma_{{\bf k}} - \mu$, the energy dispersion of 
holon 
with $\gamma_{{\bf k}}= ( \cos k_x + \cos k_y )$ and 
$ v( {\bf k}^{'} - {\bf k} ) = -J|\Delta_f|^2 \gamma_{{\bf k}} $, the momentum 
space representation of holon-holon interaction.
$N$ is the total number of sites for the two dimensional lattice of interest and 
$\beta = \frac{1}{k_B T}$, the inverse temperature.
Considering the ladder diagrams for the holon-holon scattering, we obtain the 
Bethe-Salpeter equation for the t-matrix, 
\bqa
\lefteqn{< k^{'}, -k^{'}+q | t | k, -k+q >_{U(1)}  =  
v( {\bf k}^{'} - {\bf k} ) }\nn
 &&-\frac{1}{N\beta} \sum_{k^{''}}  v( {\bf k}^{'} - {\bf k}^{''} ) g(k^{''}) g( 
-k^{''}+q) \times \nn
 && < k^{''}, -k^{''}+q | t | k, -k+q >_{U(1)},
\label{dyson_u1_app}
\eqa
where $ g(k) = <b({\bf k},\omega_n) b({\bf k},\omega_n)^\dagger>$, the holon 
Matsubara Green's function and $k = (\omega_n, k_x, 
k_y)$ is the 
three-component vector of the energy and momentum. 
Using the fact that the holon-holon interaction $v( {\bf k}^{'} - {\bf k}^{''} 
)$ is frequency independent, we sum the Matsubara 
frequency 
$k^{''}_0$,
\bqa
\frac{1}{ \beta} \sum_{k^{''}_0} g( k^{''} ) g( -k^{''} + q ) 
& = & \frac{1}{\beta} \sum_{k^{''}_0} 
\frac{1}{ ik_0^{''} - \epsilon({\bf k}^{''}) } 
\frac{1}{ i(-k_0^{''}+q_0) - \epsilon(-{\bf k}^{''} + {\bf q}) } \nn
& = &  -\frac{n(\epsilon({\bf k}^{''})) + e^{\beta \epsilon(-{\bf k}^{''}+ {\bf 
q}) } n(\epsilon(-{\bf k}^{''}+{\bf q}))}{ iq_0 - 
(\epsilon(-{\bf k}^{''}+{\bf q}) + \epsilon({\bf k}^{''})) }.
\label{k0_u1_app}
\eqa
Inserting Eq.(\ref{k0_u1_app}) into Eq.(\ref{dyson_u1_app}) 
and defining $t_{{\bf k}^{'}, {\bf k}}( {\bf q}, q_0 ) \equiv < k^{'}, -k^{'}+q 
| t | k, -k+q >_{U(1)}$,
we obtain the Bethe-Salpeter equation for the t-matrix, 
\bqa
\sum_{ {\bf k}^{''}} \left( \delta_{{\bf k}^{'}, {\bf k}^{''}} - m_{{\bf k}^{'}, 
{\bf k}^{''}} ( {\bf q}, q_0 )  \right) t_{{\bf 
k}^{''}, {\bf 
k}}( {\bf q}, q_0 ) = v( {\bf k}^{'} -  {\bf k} ),
\label{eq:matrix_equation_u1_app}
\eqa
where 
\bqa
\lefteqn{ m_{{\bf k}^{'}, {\bf k}^{''}} ( {\bf q}, q_0 ) \equiv 
-\frac{1}{N\beta} \sum_{k^{''}_0}  v( {\bf k}^{'} - {\bf k}^{''} ) 
g(k^{''}) g( 
-k^{''}+q) } \nn
&& = -\frac{ v( {\bf k}^{'} - {\bf k}^{''} )}{N\beta} \left[ \sum_{k^{''}_0} 
\frac{1}{ ik_0^{''} - \epsilon({\bf k}^{''}) } 
\frac{1}{ 
i(-k_0^{''}+q_0) - \epsilon(-{\bf k}^{''} + {\bf q}) } \right] \nn
&& = \frac{1}{N} v( {\bf k}^{'} -  {\bf k}^{''} ) \frac{n(\epsilon({\bf 
k}^{''})) + e^{\beta \epsilon(-{\bf k}^{''}+ {\bf q}) } 
n(\epsilon(-{\bf 
k}^{''}+{\bf q}))}{ iq_0 - (\epsilon(-{\bf k}^{''}+{\bf q}) + \epsilon({\bf 
k}^{''})) }
\label{eq:kernel_u1_app}
\eqa
with $n(\epsilon) = \frac{1}{e^{\beta \epsilon} - 1}$, the boson distribution 
function.

\renewcommand{\theequation}{D\arabic{equation}}
\setcounter{equation}{0}
\section*{APPENDIX D: Derivation of Eqs.(6) and (7)}
Introducing a uniform hopping order parameter, $\chi_{ji}=\chi$ and a $d$-wave 
spinon pairing order parameters, $ \Delta_{ji}^{f}= 
\pm 
\Delta_f$(the sign $+(-)$ is for the ${\bf ij}$ link parallel to $\hat x$ ($\hat 
y$)), 
we obtain the energy-momentum space representation of the action for the SU(2) 
theory,
\bqa
&&  S^b_{SU(2)}  =  \sum_{{\bf k},\omega_n} 
\left( 
b_1({\bf k}, \omega_n)^\dagger,
b_2({\bf k}, \omega_n)^\dagger 
\right)
\left( \begin{array}{cc} 
-i\omega_n -t \chi \gamma_{{\bf k}} - \mu & t \Delta_f \varphi_{{\bf k}} \\ 
t \Delta_f \varphi_{{\bf k}} & -i\omega_n + t \chi \gamma_{{\bf k}} - \mu   
\end{array} 
\right) 
\left( \begin{array}{c} 
b_1({\bf k}, \omega_n)  \\
b_2({\bf k}, \omega_n) 
\end{array} \right) \nn
&& +  \frac{1}{2N}\sum_{{\bf k}, {\bf k}^{'}, {\bf q}, \alpha, \beta } 
\sum_{\omega_n, \omega_n^{'}, \nu_n}
v( {\bf k } - {\bf k}^{'}) 
b_{\beta}(-{\bf k}^{'} + {\bf q}, -\omega_n^{'}+\nu_n)^\dagger 
b_{\alpha}({\bf k}^{'}, \omega_n^{'})^\dagger 
b_{\alpha}({\bf k}, \omega_n)
b_{\beta}(-{\bf k} + {\bf q}, -\omega_n+\nu_n), \nn
\label{eq:su2_model_momentum_app}
\eqa
where 
$b_{\alpha}({\bf k},\omega_n) = \frac{1}{\sqrt{N\beta}} \int_0^\beta d \tau 
\sum_{i} e^{i( \omega_n \tau  - {\bf k} \cdot {\bf 
r}_i)} 
b_{i\alpha}(\tau)$ and $\varphi_{{\bf k}} = ( \cos k_x- \cos k_y )$ 
with $\alpha$, $\beta$ ($=1,2$), two components of SU(2) holon. 
Considering the ladder diagrams for the holon-holon scattering, we obtain the 
Bethe-Salpeter equation for the t-matrix,
\bqa
&& < k^{'}, \alpha^{'}; -k^{'}+q, \beta^{'} | t | k,\alpha; -k+q,\beta >_{SU(2)}  
 =  v( {\bf k}^{'} - {\bf k} ) \delta_{\alpha^{'},\alpha} 
\delta_{\beta^{'},\beta} \nn
&& -\frac{1}{N\beta} \sum_{k^{''},\alpha^{''},\beta^{''}}  v( {\bf k}^{'} - {\bf 
k}^{''} ) g_{\alpha^{'}\alpha^{''}}(k^{''}) 
g_{\beta^{'}\beta^{''}}( -k^{''}+q) \nn
&& \times < k^{''}, \alpha^{''}; -k^{''}+q, \beta^{''} | t | k,\alpha; 
-k+q,\beta >_{SU(2)},
\label{dyson_su2_app}
\eqa
where $g_{\alpha \beta}(k) = <b_{\alpha} ( {\bf k}, \omega_n) b_{\beta} ( {\bf 
k}, \omega_n)^\dagger>$, the holon Matsubara Green's 
function.

To calculate the holon Matsubara Green's function in the diagonalized basis of 
the one-body term in the action 
(Eq.(\ref{eq:su2_model_momentum_app})), we introduce the unitary transformation 
for the holon field,
\bq
\left( \begin{array}{c} 
b_1({\bf k}, \omega_n )  \\
b_2({\bf k}, \omega_n )
\end{array} \right) 
= U({\bf k})
\left( \begin{array}{c} 
b^{'}_1({\bf k}, \omega_n ) \\
b^{'}_2({\bf k}, \omega_n )
\end{array} \right)
\eq
where
\bq
U_{\alpha \beta}({\bf k}) = \left( \begin{array}{cc} 
u_{{\bf k}} & -v_{{\bf k}} \\ 
v_{{\bf k}} & u_{{\bf k}} \end{array} \right)
\eq
with 
$ u_{{\bf k}} = \frac{1}{\sqrt{2}}\sqrt{1-\frac{t\chi \gamma_{{\bf k}}}{E_{{\bf 
k}}}}$, 
$ v_{{\bf k}} = \frac{sgn(\varphi_k)}{\sqrt{2}}\sqrt{1+\frac{t\chi \gamma_{{\bf 
k}}}{E_{{\bf k}}}}$,
$E({\bf k}) =  t \sqrt{ (\chi \gamma_{{\bf k}})^2 + ( \Delta_f \varphi_{{\bf 
k}})^2} $ and
$b^{'}_{\alpha}({\bf k}, \omega_n)$, the quasi-holon field.
Then the one-body action in Eq.(\ref{eq:su2_model_momentum_app}) becomes
\bqa
&&  S^{b0}_{SU(2)}  =  
\sum_{{\bf k},\omega_n} 
\left( 
b_1({\bf k}, \omega_n)^\dagger,
b_2({\bf k}, \omega_n)^\dagger 
\right)
\left( \begin{array}{cc} 
-i\omega_n -t \chi \gamma_{{\bf k}} - \mu & t \Delta_f \varphi_{{\bf k}} \\ 
t \Delta_f \varphi_{{\bf k}} & -i\omega_n + t \chi \gamma_{{\bf k}} - \mu   
\end{array} 
\right) 
\left( \begin{array}{c} 
b_1({\bf k}, \omega_n)  \\
b_2({\bf k}, \omega_n) 
\end{array} \right) \nn
&&= 
\sum_{{\bf k},\omega_n} 
\left( 
b^{'}_1({\bf k}, \omega_n)^\dagger,
b^{'}_2({\bf k}, \omega_n)^\dagger 
\right)
\left( \begin{array}{cc} 
-i\omega_n + E_1({\bf k})  & 0 \\ 
 0  & -i\omega_n + E_2({\bf k}) \end{array} 
\right) 
\left( \begin{array}{c} 
b^{'}_1({\bf k}, \omega_n)  \\
b^{'}_2({\bf k}, \omega_n) 
\end{array} \right),
\label{su2_diagonalized_action_app}
\eqa
where the quasi-holon energy is 
$E_1({\bf k}) =  t \sqrt{ (\chi \gamma_{{\bf k}})^2 + ( \Delta_f \varphi_{{\bf 
k}})^2} - \mu$ and
$E_2({\bf k}) =  -t \sqrt{ (\chi \gamma_{{\bf k}})^2 + ( \Delta_f \varphi_{{\bf 
k}})^2} - \mu$.
From the action in Eq.(\ref{su2_diagonalized_action_app}) we readily obtain the 
holon Matsubara Green's function, 
\bqa
&& g_{\alpha \beta}(k) 
= < b_{\alpha}({\bf k}, \omega_n) b_{\beta}^\dagger({\bf k}, \omega_n) > \nn
&& = U_{\alpha \alpha^{'}} ({\bf k}) 
 < b^{'}_{\alpha^{'}}({\bf k}, \omega_n) b^{' \dagger}_{\beta^{'}}({\bf k}, 
\omega_n) > 
(U({\bf k})^\dagger)_{\beta^{'} \beta} \nn
&& = U_{\alpha \alpha^{'}} ({\bf k}) 
\left( \begin{array}{cc}
-\frac{1}{ i \omega_n + E_1({\bf k}) } & 0 \\
0 & -\frac{1}{ i \omega_n + E_2({\bf k}) } 
\end{array} \right)_{\alpha^{'} \beta^{'}}
(U({\bf k})^\dagger)_{\beta^{'} \beta},
\eqa
where $E_1({\bf k})$ and $E_2({\bf k})$
are the holon quasiparticle energy in the upper and lower band respectively and
sgn($\varphi_k$) denotes the sign of $(\cos k_x - \cos k_y)$. 
It is of note that there are two energy bands owing to the two kinds of holons 
in the SU(2) theory.

Using the fact that the holon-holon interaction $v( {\bf k}^{'} - {\bf k}^{''} 
)$ is frequency independent, we sum the Matsubara 
frequency 
$k^{''}_0$ in Eq.(\ref{dyson_su2_app}),
\bqa
&& \frac{1}{\beta} \sum_{k^{''}_0} g_{\alpha^{'} \alpha^{''}}( k^{''} ) 
g_{\beta^{'} \beta^{''}}( -k^{''} + q ) \nn
&& =  
\frac{1}{\beta} \sum_{k^{''}_0} \sum_{\alpha_1,\beta_1}
\left(
U({\bf k})_{\alpha^{'} \alpha_1} 
\frac{1}{ ik_0^{''} - E_{\alpha_1}({\bf k}^{''}) } 
U({\bf k})^\dagger_{\alpha_1 \alpha^{''}} 
\right) \nn
&& \times 
\left(
U(-{\bf k}+{\bf q})_{\beta^{'} \beta_1} 
\frac{1}{ i(-k_0^{''}+q_0) - E_{\beta_1}(-{\bf k}^{''} + {\bf q}) } 
U^\dagger(-{\bf k}+{\bf q})_{\beta_1 \beta^{''}} 
\right) \nn
& = &  
-U({\bf k})_{\alpha^{'} \alpha_1} 
U({\bf k})^\dagger_{\alpha_1 \alpha^{''}} 
U(-{\bf k}+{\bf q})_{\beta^{'} \beta_1} 
U^\dagger(-{\bf k}+{\bf q})_{\beta_1 \beta^{''}}  \nn
&& \times 
\frac{n(E_{\alpha_1}({\bf k}^{''})) + e^{\beta E_{\beta_1}(-{\bf k}^{''}+ {\bf 
q}) } n(E_{\beta_1}(-{\bf k}^{''}+{\bf q}))}{ iq_0 - 
(E_{\beta_1}(-{\bf k}^{''}+{\bf q}) + E_{\alpha_1}({\bf k}^{''})) }.
\label{k0_su2_app}
\eqa
Inserting Eq.(\ref{k0_su2_app}) into Eq.(\ref{dyson_su2_app}) 
and defining $t^{\alpha^{'} \beta^{'} \alpha \beta}_{{\bf k}^{'}, {\bf k}}( {\bf 
q}, q_0 ) \equiv
< k^{'}, \alpha^{'}; -k^{'}+q, \beta^{'} | t | k,\alpha; -k+q,\beta >_{SU(2)}$,
we obtain the Bethe-Salpeter equation for the t-matrix, 
\bqa
\lefteqn{\sum_{ {\bf k}^{''}, \alpha^{''}, \beta^{''} } \left( \delta_{{\bf 
k}^{'}, {\bf k}^{''}} \delta_{\alpha^{'} \alpha^{''}} 
\delta_{\beta^{'} \beta^{''}} - m^{\alpha^{'} \beta^{'} \alpha^{''} \beta^{''} 
}_{{\bf k}^{'}, {\bf k}^{''}} ( {\bf q}, q_0 )  
\right) t^{ 
\alpha^{''} \beta^{''} \alpha \beta }_{{\bf k}^{''}, {\bf k}}( {\bf q}, q_0 ) 
}\nn
&& = v( {\bf k}^{'} -  {\bf k} ) \delta_{\alpha^{'} \alpha} \delta_{\beta^{'} 
\beta}, \hspace{4cm}
\label{eq:matrix_equation_su2_app}
\eqa
where 
\bqa
&& m^{\alpha^{'} \beta^{'} \alpha^{''} \beta^{''} }_{{\bf k}^{'}, {\bf k}^{''}} 
( {\bf q}, q_0 )  = 
\frac{1}{N} \sum_{\alpha^{'}_1 \beta^{'}_1} v( {\bf k}^{'} -  {\bf k}^{''} ) 
\frac{n(E_{\alpha^{'}_1}({\bf k}^{''})) + e^{\beta 
E_{\beta^{'}_1}(-{\bf k}^{''}+{\bf q})} n(E_{\beta^{'}_1}(-{\bf k}^{''}+{\bf 
q}))}{ iq_0 - (E_{\alpha^{'}_1}({\bf k}^{''}) + 
E_{\beta^{'}_1}(-{\bf k}^{''}+ {\bf q} )) } \times  \nn
&& U_{\alpha^{'}\alpha^{'}_1}({\bf k}^{''}) U_{\beta^{'}\beta^{'}_1}(- {\bf 
k}^{''} + {\bf q}) U^\dagger_{\alpha^{'}_1 
\alpha^{''}}({\bf 
k}^{''}) U^\dagger_{\beta^{'}_1 \beta^{''} }(- {\bf k}^{''} + {\bf q}). 
\label{eq:kernel_su2_app}
\eqa

\newpage
\centerline{TABLE CAPTIONS}
\begin{itemize}

\item[Table 1 ]
Weights of the $s$-, $p_x$-, $p_y$- and $d$-wave contributions to the holon 
pairing order parameter corresponding to the lowest 
energy in the 
U(1) holon-pair boson theory.

\item[Table 2 ]
Weights of the $s$-, $p_x$-, $p_y$- and $d$-wave contributions in each 
$b_{\alpha}$-$b_{\beta}$ scattering channel with 
$\alpha,\beta=1, 2$  to 
the first holon pair order parameter corresponding to the lowest energy in the 
SU(2) holon-pair boson theory.

\item[Table 3 ]
Weights of the $s$-, $p_x$-, $p_y$- and $d$-wave contributions in each 
$b_{\alpha}$-$b_{\beta}$ scattering channel with 
$\alpha,\beta=1, 2$ to 
the second holon pair order parameter corrsponding to the lowest energy in the 
SU(2) holon-pair boson theory.

\item[Table 4 ]
The orbital state and energy at various hole doping $x$ and temperatures ($T/t = 
T_c/t + 0.001$) for underdoping ($x=0.04$,  
$T_c/t=0.034$), 
optimal doping($x=0.07$, $T_c/t=0.044$) and overdoping ($x=0.1$, $T_c/t=0.041$) 
respectively with the U(1) holon-pair boson theory.
The calculations are done on the $N= 10 \times 10$ lattice with the use of 
$J/t=0.3$.
Here, ${\bf s}$ denotes $(\cos k_x + \cos k_y)$; ${\bf d}$, $(\cos k_x - \cos 
k_y)$; ${\bf p_x}$, $\sin k_x$ and ${\bf p_y}$, $\sin 
k_y$.

\item[Table 5 ]
The orbital and energy at temperatures $T/t = T_c/t +0.001$ for underdoping 
($x=0.07$, $T_c/t=0.027$), optimal doping($x=0.13$, 
$T_c/t=0.037$) 
and overdoping ($x=0.19$, $T_c/t=0.031$) respectively in the SU(2) holon-pair 
boson theory.
\end{itemize}

\newpage
\begin{table}[h]
\begin{center}
\begin{tabular}{|c|c|c|c|}
	weight & $x=0.04$  	& $x=0.07$      & $x=0.1$  \\
	\hline
	\hline
	$a_{s}$   & ${\bf 1}$     	&  ${\bf 1}$	&  ${\bf 1}$   \\			  
	\hline
	$a_{p_x}$ & $-1.8 \times 10^{-14}$  &  $-2.5 \times 10^{-14}$	&  $-5.3 
\times 10^{-15}$   \\			  
	\hline
	$a_{p_y}$ & $-1.4 \times 10^{-14}$  &  $-1.8 \times 10^{-14}$	&  $-3.4 
\times 10^{-14}$   \\			  
	\hline
	$a_{d}$   & $-1.3 \times 10^{-15}$  &  $-8.9 \times 10^{-15}$	&  $-3.2 
\times 10^{-15}$   \\			  
\end{tabular}
\end{center}
\caption{}
\end{table}

\newpage
\begin{table}[h]
\begin{center}
\begin{tabular}{|c|c|c|c|}
	weight & $x=0.07$  	& $x=0.13$      & $x=0.19$  \\
	\hline
	\hline
 $a_{s,11}$ & $ {\bf 1} $  & $ {\bf 1}  $ & $ {\bf 1} $   \\ 
 \hline
 $a_{p_x,11}$ & $ -5.7 \times 10^{-15} $  & $ -1.3 \times 10^{-20}  $ & $ 6.1 
\times 10^{-21} $   \\ 
 \hline
 $a_{p_y,11}$ & $ -2.4 \times 10^{-14} $  & $ 1.2 \times 10^{-21}  $ & $ -1.3 
\times 10^{-21} $   \\ 
 \hline
 $a_{d,11}$ & $ -1.3 \times 10^{-15} $  & $ -3.5 \times 10^{-27}  $ & $ -2.9 
\times 10^{-21} $   \\ 
 \hline
 $a_{s,12}$ & $ 9.2 \times 10^{-20} $  & $ 1.8 \times 10^{-22}  $ & $ 4.9 \times 
10^{-18} $   \\ 
 \hline
 $a_{p_x,12}$ & $ -6.7 \times 10^{-27} $  & $ 1.2 \times 10^{-33}  $ & $ 6.2 
\times 10^{-25} $   \\ 
 \hline
 $a_{p_y,12}$ & $ 2.2 \times 10^{-27} $  & $ 6.6 \times 10^{-28}  $ & $ -5.9 
\times 10^{-31} $   \\ 
 \hline
 $a_{d,12}$ & $ -4.3 \times 10^{-26} $  & $ -2.2 \times 10^{-33}  $ & $ -3.5 
\times 10^{-31} $   \\ 
 \hline
 $a_{s,21}$ & $ 5.3 \times 10^{-22} $  & $ 1.8 \times 10^{-22}  $ & $ 4.9 \times 
10^{-18} $   \\ 
 \hline
 $a_{p_x,21}$ & $ -9.3 \times 10^{-29} $  & $ 1.2 \times 10^{-33}  $ & $ 6.1 
\times 10^{-25} $   \\ 
 \hline
 $a_{p_y,21}$ & $ 3.1 \times 10^{-29} $  & $ 6.6 \times 10^{-28}  $ & $ -5.8 
\times 10^{-31} $   \\ 
 \hline
 $a_{d,21}$ & $ -2.5 \times 10^{-28} $  & $ 1.4 \times 10^{-29}  $ & $ 1.4 
\times 10^{-29} $   \\ 
 \hline
 $a_{s,22}$ & $ {\bf 1} $  & $ {\bf 1}  $ & $ {\bf 1} $   \\ 
 \hline
 $a_{p_x,22}$ & $ 1.0 \times 10^{-20} $  & $ 7.2 \times 10^{-21}  $ & $ -1.1 
\times 10^{-20} $   \\ 
 \hline
 $a_{p_y,22}$ & $ 1.2 \times 10^{-20} $  & $ -5.2 \times 10^{-22}  $ & $ -6.2 
\times 10^{-21} $   \\ 
 \hline
 $a_{d,22}$ & $ 7.7 \times 10^{-21} $  & $ -2.0 \times 10^{-28}  $ & $ 1.6 
\times 10^{-27} $   \\ 
\end{tabular}
\end{center}
\caption{}
\end{table}

\begin{table}[h]
\begin{center}
\begin{tabular}{|c|c|c|c|}
	weight & $x=0.07$  	& $x=0.13$      & $x=0.19$  \\
	\hline
	\hline
 $a_{s,11}$ & $ {\bf 1} $  & $ {\bf 1}  $ & $ {\bf 1} $   \\ 
 \hline
 $a_{p_x,11}$ & $ -4.0 \times 10^{-15} $  & $ -7.2 \times 10^{-21}  $ & $ 1.1 
\times 10^{-20} $   \\ 
 \hline
 $a_{p_y,11}$ & $ -4.4 \times 10^{-15} $  & $ 5.2 \times 10^{-22}  $ & $ 6.2 
\times 10^{-21} $   \\ 
 \hline
 $a_{d,11}$ & $ -1.5 \times 10^{-14} $  & $ 1.9 \times 10^{-28}  $ & $ -1.6 
\times 10^{-27} $   \\ 
 \hline
 $a_{s,12}$ & $ -4.9 \times 10^{-14} $  & $ 1.6 \times 10^{-15}  $ & $ 6.8 
\times 10^{-22} $   \\ 
 \hline
 $a_{p_x,12}$ & $ -2.7 \times 10^{-14} $  & $ 2.3 \times 10^{-15}  $ & $ 3.2 
\times 10^{-27} $   \\ 
 \hline
 $a_{p_y,12}$ & $ -4.9 \times 10^{-14} $  & $ 1.6 \times 10^{-22}  $ & $ -3.7 
\times 10^{-34} $   \\ 
 \hline
 $a_{d,12}$ & $ {\bf -1.3 \times 10^{-4}} $  & $ {\bf -8.3 \times 10^{-6}}  $ & 
$ -1.4 \times 10^{-33} $   \\ 
 \hline
 $a_{s,21}$ & $ -4.7 \times 10^{-14} $  & $ 7.5 \times 10^{-11}  $ & $ 6.8 
\times 10^{-22} $   \\ 
 \hline
 $a_{p_x,21}$ & $ 2.6 \times 10^{-14} $  & $ -1.9 \times 10^{-10}  $ & $ 3.2 
\times 10^{-27} $   \\ 
 \hline
 $a_{p_y,21}$ & $ 1.7 \times 10^{-14} $  & $ 2.3 \times 10^{-23}  $ & $ -3.7 
\times 10^{-34} $   \\ 
 \hline
 $a_{d,21}$ & $ {\bf -1.3 \times 10^{-4}} $  & $ {\bf -8.3 \times 10^{-6}}  $ & 
$ -1.4 \times 10^{-29} $   \\ 
 \hline
 $a_{s,22}$ & $ {\bf -1} $  & $ {\bf -1}  $ & $ {\bf -1} $   \\ 
 \hline
 $a_{p_x,22}$ & $ 7.1 \times 10^{-15} $  & $ 2.5 \times 10^{-14}  $ & $ 6.3 
\times 10^{-21} $   \\ 
 \hline
 $a_{p_y,22}$ & $ 3.1 \times 10^{-14} $  & $ 1.2 \times 10^{-20}  $ & $ 2.7 
\times 10^{-21} $   \\ 
 \hline
 $a_{d,22}$ & $ 2.5 \times 10^{-15} $  & $ 1.0 \times 10^{-14}  $ & $ -1.9 
\times 10^{-27} $   \\ 
\end{tabular}
\end{center}
\caption{}
\end{table}

\begin{table}[h]
\begin{center}
\begin{tabular}{|c|c|c|c|}
	eigenvector & $\omega/t$ & $\omega/t$ & $\omega/t$ \\
		    &  ($x=0.04$)  & ($x=0.07$)  &  ($x=0.1$)  \\
	\hline
	\hline
	${\bf s}$		& $0.0111$  &  $0.0089$	&  $0.0063$   \\			  
	\hline
	\hline
	${\bf s}$		& $0.7221$  &  $1.1180$	&  $1.5592$   \\			  
	\hline
	${\bf p_x}$, ${\bf p_y }$	& $0.7236$  &  $1.1186$	&  $1.5594$   \\			  
	\hline
	${\bf d}$		& $0.7238$  &  $1.1187$	&  $1.5595$   \\			  
	\hline
	\hline
	${\bf s}$		& $1.4306$  &  $2.2249$	&  $3.1108$   \\			  
	\hline
	${\bf p_x}$, ${\bf p_y }$	& $1.4316$  &  $2.2252$	&  $3.1109$   \\			  
	\hline
	\hline
	${\bf s}$		& $2.5769$  &  $4.0157$	&  $5.6212$   \\			  
	\hline
	${\bf p_x}$, ${\bf p_y }$	& $2.5773$  & $4.01585$ & $5.62126$   \\			  
	\hline
	${\bf d}$		& $2.5775$  & $4.01594$ & $5.62129$ 
\end{tabular}
\end{center}
\caption{}
\end{table}

\begin{table}[h]
\begin{center}
\begin{tabular}{|c|c|c|c|}
	eigenvector & $\omega/t$ & $\omega/t$ & $\omega/t$ \\
		& ($x=0.07$)  & ($x=0.13$)  & ($x=0.19$)  \\
	\hline
	\hline
	${\bf 
s}(\delta_{\alpha,1}\delta_{\beta,1}+\delta_{\alpha,2}\delta_{\beta,2}),$	
	& $0.0097$  &  $0.0080$	&  $0.0054$   \\			  
	${\bf 
s}(\delta_{\alpha,1}\delta_{\beta,1}-\delta_{\alpha,2}\delta_{\beta,2}) 
	-a{\bf 
d}(\delta_{\alpha,1}\delta_{\beta,2}+\delta_{\alpha,2}\delta_{\beta,1})$	
	& ($a=1.3\times 10^{-4}$)	& ($a=8.3\times10^{-6}$)	& 
($a<10^{-10}$)\\
	\hline
	\hline
	${\bf 
s}(\delta_{\alpha,1}\delta_{\beta,1}+\delta_{\alpha,2}\delta_{\beta,2}),$	
	& $0.3494$  &  $0.5899$	&  $0.8935$   \\			  
	${\bf 
s}(\delta_{\alpha,1}\delta_{\beta,1}-\delta_{\alpha,2}\delta_{\beta,2}) 
	-a{\bf 
d}(\delta_{\alpha,1}\delta_{\beta,2}+\delta_{\alpha,2}\delta_{\beta,1})$	
	& ($a=4\times 10^{-3}$)	& ($a=1.4\times10^{-3}$)	& 
($a=4.4\times10^{-4}$)\\
	\hline
	${\bf 
p_x}(\delta_{\alpha,1}\delta_{\beta,1}+\delta_{\alpha,2}\delta_{\beta,2}),$	
	& $0.3510$  &  $0.5905$	&  $0.8936$   \\			  
	${\bf p_y 
}(\delta_{\alpha,1}\delta_{\beta,1}+\delta_{\alpha,2}\delta_{\beta,2}),$	
	& ($a=3.6\times 10^{-2}$)	& ($a=1.3\times10^{-2}$)	& 
($a=4.1\times10^{-3}$)\\
	${\bf 
p_x}(\delta_{\alpha,1}\delta_{\beta,1}-\delta_{\alpha,2}\delta_{\beta,2}) 
	+a{\bf 
p_x}(\delta_{\alpha,1}\delta_{\beta,2}+\delta_{\alpha,2}\delta_{\beta,1}),$	
	&   &  	&     \\			  
	${\bf p_y 
}(\delta_{\alpha,1}\delta_{\beta,1}-\delta_{\alpha,2}\delta_{\beta,2}) 
	-a{\bf p_y 
}(\delta_{\alpha,1}\delta_{\beta,2}+\delta_{\alpha,2}\delta_{\beta,1})$	
	&   &  	&     \\			  
	\hline
	${\bf 
d}(\delta_{\alpha,1}\delta_{\beta,1}+\delta_{\alpha,2}\delta_{\beta,2}),$	
	& $0.3512$  &  $0.5906$	&  $0.8997$   \\			  
	${\bf 
d}(\delta_{\alpha,1}\delta_{\beta,1}-\delta_{\alpha,2}\delta_{\beta,2}) 
	-a{\bf 
s}(\delta_{\alpha,1}\delta_{\beta,2}+\delta_{\alpha,2}\delta_{\beta,1})$	
	& ($a=3.4\times 10^{-1}$)	& ($a=1.2\times10^{-1}$)	& 
($a=3.9\times10^{-2}$)\\
	\hline
	\hline
	${\bf 
s}(\delta_{\alpha,1}\delta_{\beta,1}+\delta_{\alpha,2}\delta_{\beta,2}),$	
	& $0.6910$  &  $1.1707$	&  $1.7809$   \\			  
	${\bf 
s}(\delta_{\alpha,1}\delta_{\beta,1}-\delta_{\alpha,2}\delta_{\beta,2}) 
	-a{\bf 
d}(\delta_{\alpha,1}\delta_{\beta,2}+\delta_{\alpha,2}\delta_{\beta,1})$	
	& ($a=2.5\times 10^{-4}$)	& ($a=1.8\times10^{-5}$)	& 
($a<10^{-6}$)\\
	\hline
	${\bf 
p_x}(\delta_{\alpha,1}\delta_{\beta,1}+\delta_{\alpha,2}\delta_{\beta,2}),$	
	& $0.6920$  &  $1.1711$	&  $1.7810$   \\			  
	${\bf p_y 
}(\delta_{\alpha,1}\delta_{\beta,1}+\delta_{\alpha,2}\delta_{\beta,2}),$	
	& ($a=2.6\times 10^{-4}$)	& ($a=2.0\times10^{-5}$)	& 
($a<10^{-6}$)\\
	${\bf 
p_x}(\delta_{\alpha,1}\delta_{\beta,1}-\delta_{\alpha,2}\delta_{\beta,2}) 
	+a{\bf 
p_x}(\delta_{\alpha,1}\delta_{\beta,2}+\delta_{\alpha,2}\delta_{\beta,1}),$	
	&   &  	&     \\			  
	${\bf p_y 
}(\delta_{\alpha,1}\delta_{\beta,1}-\delta_{\alpha,2}\delta_{\beta,2}) 
	-a{\bf p_y 
}(\delta_{\alpha,1}\delta_{\beta,2}+\delta_{\alpha,2}\delta_{\beta,1})$	
	&   &  	&     \\			  
	\hline
	\hline
	${\bf 
s}(\delta_{\alpha,1}\delta_{\beta,1}+\delta_{\alpha,2}\delta_{\beta,2}),$	
	& $1.2030$  &  $2.1015$	&  $3.2154$   \\			  
	${\bf 
s}(\delta_{\alpha,1}\delta_{\beta,1}-\delta_{\alpha,2}\delta_{\beta,2}) 
	-a{\bf 
d}(\delta_{\alpha,1}\delta_{\beta,2}+\delta_{\alpha,2}\delta_{\beta,1})$	
	& ($a=9.7\times 10^{-2}$)	& ($a=3.4\times10^{-2}$)	& 
($a=1.1\times10^{-2}$)\\
	\hline
	${\bf 
p_x}(\delta_{\alpha,1}\delta_{\beta,1}+\delta_{\alpha,2}\delta_{\beta,2}),$	
	& $1.1203$  &  $2.1017$	&  $3.2155$   \\			  
	${\bf p_y 
}(\delta_{\alpha,1}\delta_{\beta,1}+\delta_{\alpha,2}\delta_{\beta,2}),$	
	& ($a=1.7\times 10^{-1}$)	& ($a=6.4\times10^{-2}$)	& 
($a=2.1\times10^{-2}$)\\
	${\bf 
p_x}(\delta_{\alpha,1}\delta_{\beta,1}-\delta_{\alpha,2}\delta_{\beta,2}) 
	+a{\bf 
p_x}(\delta_{\alpha,1}\delta_{\beta,2}+\delta_{\alpha,2}\delta_{\beta,1}),$	
	&   &  	&     \\			  
	${\bf p_y 
}(\delta_{\alpha,1}\delta_{\beta,1}-\delta_{\alpha,2}\delta_{\beta,2}) 
	-a{\bf p_y 
}(\delta_{\alpha,1}\delta_{\beta,2}+\delta_{\alpha,2}\delta_{\beta,1})$	
	&   &  	&     \\			  
	\hline
	${\bf 
d}(\delta_{\alpha,1}\delta_{\beta,1}+\delta_{\alpha,2}\delta_{\beta,2}),$	
	& $1.1236$  &  $2.1018$	&  $3.2155$   \\			  
	${\bf 
d}(\delta_{\alpha,1}\delta_{\beta,1}-\delta_{\alpha,2}\delta_{\beta,2}) 
	-a{\bf 
s}(\delta_{\alpha,1}\delta_{\beta,2}+\delta_{\alpha,2}\delta_{\beta,1})$	
	& ($a=3.4\times 10^{-1}$)	& ($a=1.2\times10^{-1}$)	& 
($a=3.9\times10^{-2}$)
\end{tabular}
\end{center}
\caption{}
\end{table}

\newpage
\centerline{FIGURE CAPTIONS}
\begin{itemize}

\item[Fig. 1 ]
The schematic spectrum of the holon pairing state at temperature (a) above $T_c$ 
and (b) below $T_c$.

\item[Fig. 2 ]
The lowest order t-matrix  starting from an l-th orbital state $w_l(k)$ to an 
intermediate state of momenta ${\bf k}_1$ and $-{\bf 
k}_1$.

\item[Fig. 3]
The treated t-matrix with the inclusion of exchange contribution.

\end{itemize}

\newpage
\begin{figure}
	\epsfxsize=12cm
	\epsfysize=10cm
	\epsffile{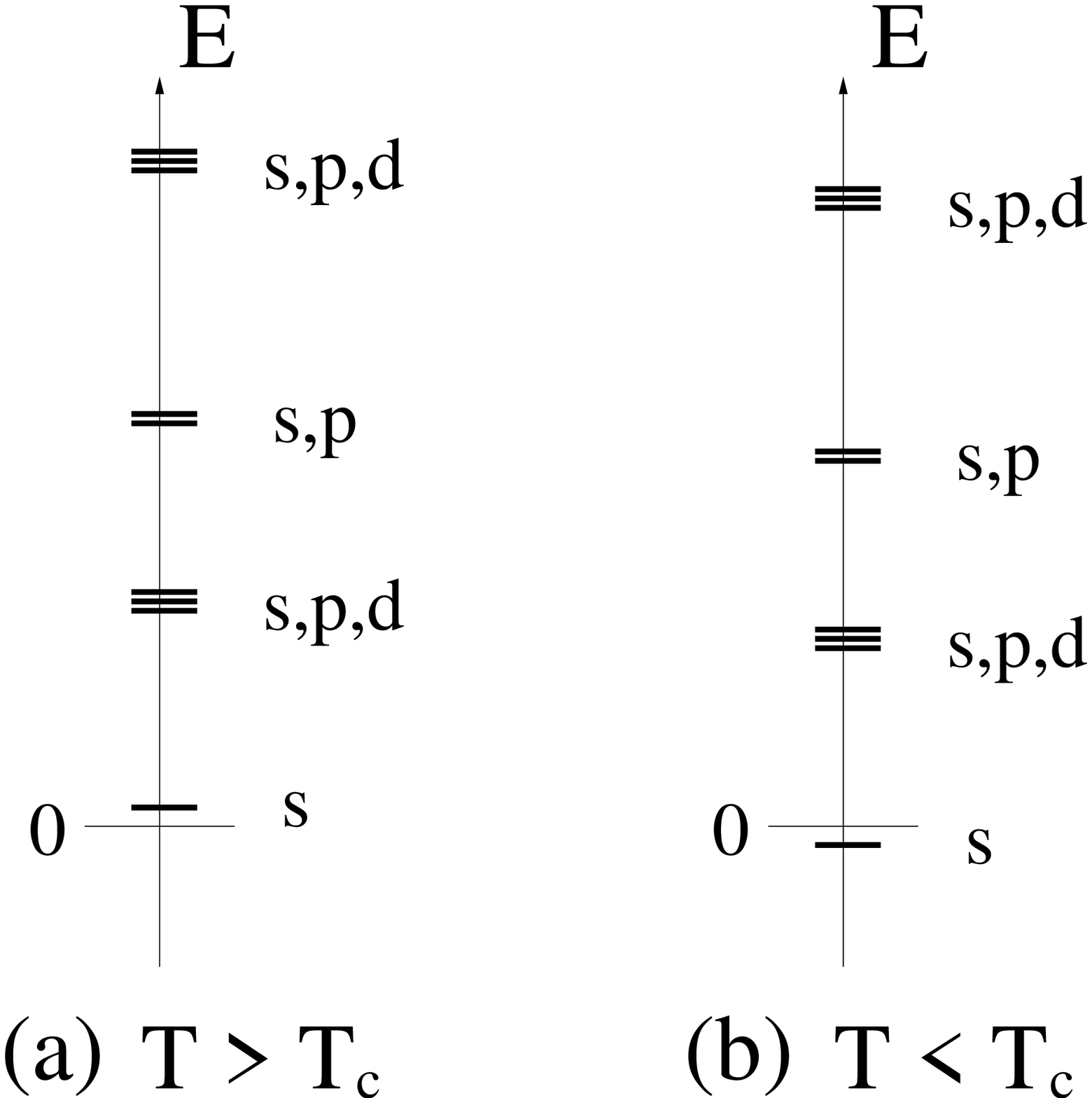}
\label{fig:1}
\caption{}
\end{figure}

\newpage
\begin{figure}
	\epsfxsize=10cm
	\epsfysize=5cm
	\epsffile{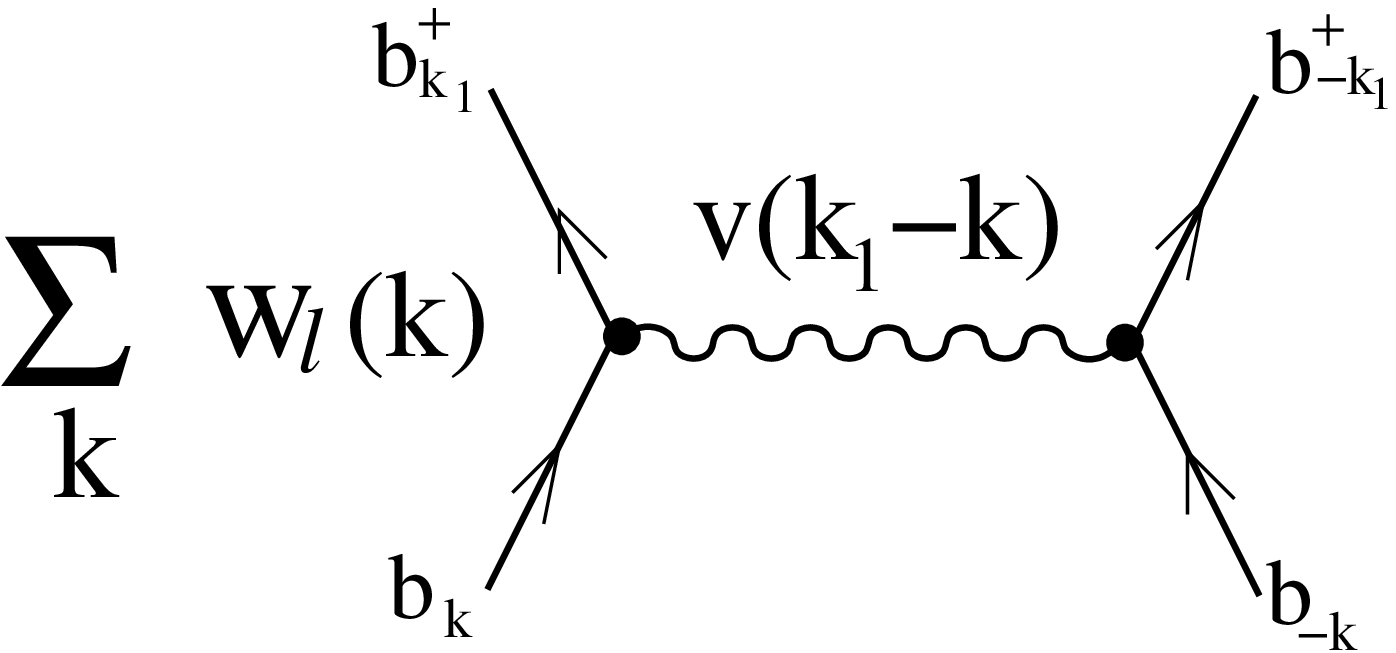}
\label{fig:2}
\caption{}
\end{figure}

\begin{figure}
	\epsfxsize=12cm
	\epsfysize=8cm
	\epsffile{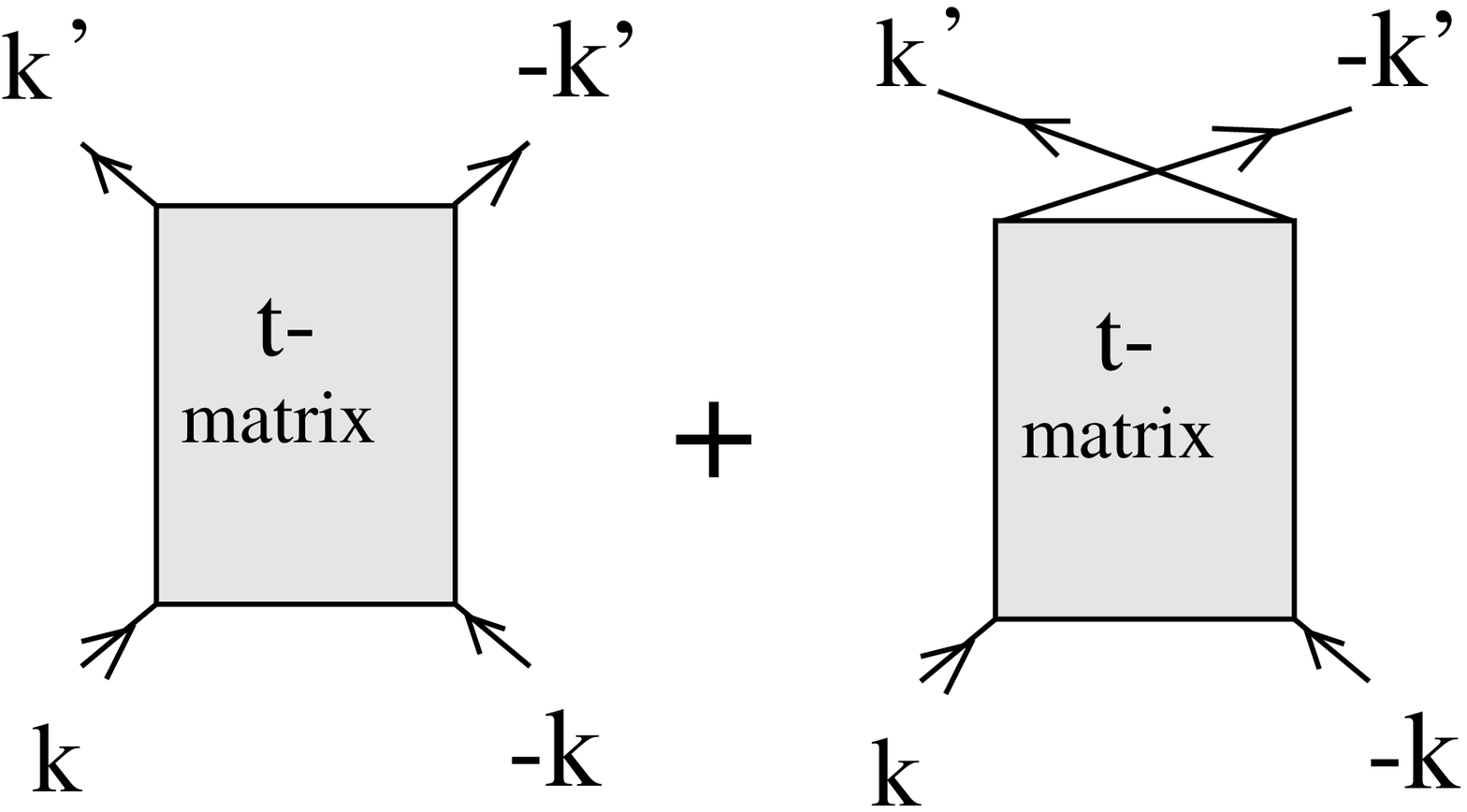}
\label{fig:3}
\caption{}
\end{figure}

%\end{multicols}
\end{document}